\newif \ifusenix
\newif \ifacm
\newif \ifmcom
\newif \ifimwut
\newif \ifflex
\newif \ifsigcomm

\sigcommfalse
\usenixfalse
\acmfalse
\mcomtrue
\imwutfalse
\flexfalse

\newcommand{\name}{GreenMO\xspace}

\ifusenix
	\documentclass[letterpaper,twocolumn,10pt]{article}
	\usepackage{usenix}
	\usepackage{times}
\fi

\ifacm
\documentclass[sigconf,10pt]{acmart}

\usepackage[english]{babel}
\renewcommand\footnotetextcopyrightpermission[1]{} 
\setcopyright{none}

\fi

\ifimwut
\documentclass[acmlarge]{acmart}

\usepackage[english]{babel}

\renewcommand\footnotetextcopyrightpermission[1]{} 

\fi

\ifmcom
 \documentclass[sigconf,10pt]{acmart}
\renewcommand\footnotetextcopyrightpermission[1]{} 
\setcopyright{none}

\fi

\usepackage{xcolor}
\usepackage{amsfonts}
\PassOptionsToPackage{hyphens}{url}\usepackage{hyperref}
\usepackage{color}
\usepackage{graphics}
\usepackage{graphicx}
\usepackage{url}
\usepackage{listings}
\usepackage{multicol}
\usepackage{multirow}
\usepackage[scaled]{helvet}
\usepackage{rotating}
\usepackage{xspace}
\usepackage{float}
\usepackage{comment}
\usepackage{enumitem}
\usepackage{amsmath}
\usepackage{mathrsfs}
\usepackage{cancel}
\usepackage{cleveref}
\usepackage[font=scriptsize,labelfont=bf]{caption}
\usepackage{tabularx}
\usepackage{pgfplots}
\usepackage{algorithm}
\usepackage{algorithmic}
\usepackage{mathtools}
\usepackage{blkarray, bigstrut} %
\usepackage{changepage} 
\pgfplotsset{compat=newest}
\pgfplotsset{plot coordinates/math parser=false} 

\newlength\figureheight
\newlength\figurewidth
\usepackage[explicit,noindentafter]{titlesec}
\titlespacing\section{4pt}{4pt plus 2pt minus 2pt}{1pt plus 2pt minus 2pt}
\titlespacing\subsection{4pt}{4pt plus 2pt minus 2pt}{1pt plus 2pt minus 2pt}
\titlespacing\subsubsection{4pt}{4pt plus 2pt minus 2pt}{1pt plus 2pt minus 2pt}

\pgfplotsset{
compat=1.11,
legend image code/.code={
\draw[mark repeat=2,mark phase=2]
plot coordinates {
(0cm,0cm)
(0.15cm,0cm)        
(0.3cm,0cm)         
};%
}
}
\usepackage[skip=0pt,font=scriptsize, labelfont=bf]{subcaption}
\usepackage{pgfplotstable}

\makeatletter

\makeatletter
\def\@maketitle{\newpage
 \null
 \setbox\@acmtitlebox\vbox{%
   \begin{center}
    {\ttlfnt \@title\par}       
{\subttlfnt \the\subtitletext\par}\vskip 1.25em
    {\baselineskip 16pt\aufnt   
     \begin{tabular}[t]{c}\@author
     \end{tabular}\par}
   \end{center}}
 \dimen0=\ht\@acmtitlebox
 \unvbox\@acmtitlebox
 \ifdim\dimen0<0.0pt\relax\vskip-\dimen0\fi}
\makeatother

\newcommand{\mathleft}{\@fleqntrue\@mathmargin\parindent}
\newcommand{\mathcenter}{\@fleqnfalse}
\makeatother

\author{Agrim Gupta$^\dagger$, Sajjad Nassirpour$^\S$, Manideep Dunna$^\dagger$, Eamon Patamasing$^\dagger$,\\ Alireza Vahid$^\S$, Dinesh Bharadia$^\dagger$}
\affiliation{
    \institution{$^\dagger$University of California San Diego, $^\S$ University of Colorado Denver}
    \country{USA}
}
\renewcommand{\shortauthors}{A. Gupta, S. Nassirpour, M. Dunna, E. Patamasing, A. Vahid, D. Bharadia}
\email{{agg003,mdunna,epatamas,dineshb}@ucsd.edu, {sajjad.nassirpour,alireza.vahid}@ucdenver.edu}
\newcommand{\authorsalt}{Raghav Subbaraman, Yeswanth Guntupalli, Shruti Jain, Rohit Kumar, Krishna Chintalapudi, Dinesh Bharadia}

\acmConference[(To appear) ACM MobiCom '23]{}{}{}

\begin{document}
\sloppy
\title{\name: Virtualized User-proportionate MIMO}



\ifimwut
\begin{abstract}

With the turn of new decade, wireless communications face a major challenge on connecting many more new users and devices, at the same time being energy efficient and minimizing its carbon footprint.
However, the current approaches to address the growing number of users and spectrum demands, like traditional fully digital architectures for Massive MIMO, demand exorbitant energy consumption.
The reason is that traditionally MIMO requires a separate RF chain per antenna, so the power consumption scales with number of antennas, instead of number of users, hence becomes energy inefficient.
Instead, \name creates a new massive MIMO architecture which is able to use many more antennas while keeping power consumption to user-proportionate numbers.
To achieve this \name introduces for the first time, the concept of virtualization of the RF chain hardware .
Instead of laying the RF chains physically to each antenna, GreenMO creates these RF chains virtually in digital domain.
This also enables \name to be the first flexible massive MIMO architecture.
Since \name's virtual RF chains are created on the fly digitally, it can tune the number of these virtual chains according to the user load, hence always flexibly consume user-proportionate power.
Thus, GreenMO paves the way for green\& flexible massive MIMO.
We prototype GreenMO on a PCB with eight antennas and evaluate it with a WARPv3 SDR platform in an office environment. 
The results demonstrate that GreenMO is $3\times$ more power-efficient than traditional Massive MIMO and $4\times$ more spectrum-efficient than traditional OFDMA systems, while multiplexing 4 users, and can save upto $40$\% power in modern 5G NR base stations. 
\end{abstract}

\ccsdesc[500]{Hardware~PCB design and layout}
\ccsdesc[300]{Human-centered computing~Mobile computing}
\ccsdesc[300]{Human-centered computing~Ubiquitous computing}
\ccsdesc[500]{Networks~Location based services}

\fi

\maketitle


\section{Introduction}

Over the past decade, wireless networks have grown exponentially, and thereof have accrued humongous carbon footprint, with the net carbon emissions rivalling that of aviation sector\cite{footprint1,footprint2}.
Extensive case studies have highlighted the need for wireless networks to `\textbf{grow sustainably}' \cite{mck,bcg}, and even consumer sentiment highlight this, with more than 70\% consumers willing to switch to a greener alternative\cite{consumer1,consumer2,consumer3}. For wireless networks, the above translates to achieving efficient use of licensed spectrum and energy at disposal.
Packing more bits per unit spectrum would support the growing data-rate by achieving spectrally efficient systems.
However, in communicating these more number of bits, the energy consumed should grow in a sub-linear manner to make wireless networks energy efficient. 
Ideally, wireless networks should achieve both spectral \& energy efficiency (SE, EE) and should be flexible, as number of users ($N$) decrease, energy consumption should proportionally scale down.


\begin{figure}[t]
    \centering
    \includegraphics[width=0.48\textwidth]{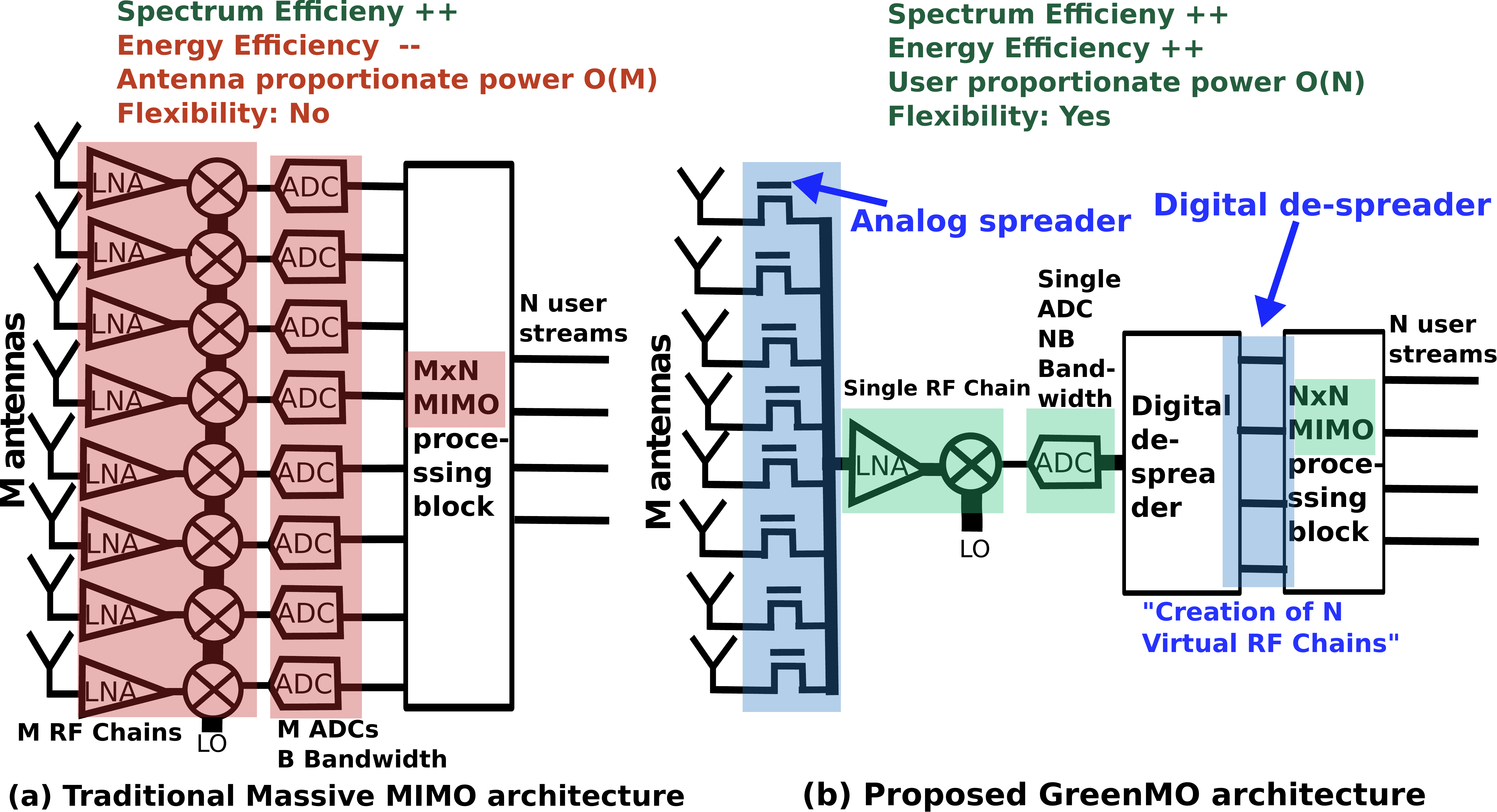}
    \caption{\name solves for energy efficiency by creating a new massive MIMO architecture with flexible user-proportionate power consumption}
    \label{fig:intro_arch}
\end{figure}


Existing solutions for scaling wireless networks largely address spectrum and energy efficiencies in isolation, and are rigid\cite{han2020energy}. 
A energy-proportionate approach to scale wireless networks is to add more spectrum bands such that the users communicate at different frequencies in an interference-free manner (referred to as frequency multiplexing). Let's say each user is provided $B$ bandwidth, and therefore $N$ users consume a total of $NB$ spectrum. 
To support this $NB$ spectrum, the energy consumption scales by $N\times$ compared to supporting $B$ bandwidth. The extra energy is spent to support receiving/transmitting the $NB$ bandwidth signal\footnote{ADC/DAC energy consumption scale linearly with sampling rate \cite{barati2020energy,yan2019performance,AD9963}}, resulting in linear scaling of energy proportionate to data bits ($N$).
However, frequency multiplexing requires new spectrum band for each new user, which is an expensive and cumbersome process, 
but it achieves user/throughput proportionate energy scaling.
It also achieves flexibility since if the number of user drops to $K < N$, the transceiver could just sample $KB$ bandwidth and therefore, would scale down the energy consumption as well. 
Hence the net sampling bandwidth of frequency multiplexing offers a simple design knob which can be adjusted in accordance to user load.


An alternate to frequency multiplexing is massive MIMO (mMIMO), which creates spatial multiplexing.
mMIMO creates $N$ distinct spatial streams one for each user by performing linear combination of the massive number of antennas signals $M$, to serve $N$ users over the same spectrum. 
Thus, mMIMO has a spectrum-cost of fixed $B$ bandwidth, but packs in $N$ times more bits, improving spectral efficiency by $N\times$ compared to frequency multiplexing.
As shown in Fig. \ref{fig:intro_arch} (a), this is achieved by digitally interfacing the $M>>N$ number of antennas, which creates the $N$ spatial streams in digital domain (with mMIMO processing). Each antenna has its own RF interfacing hardware (for up/downconversion, amplification) referred to as a RF chain, followed by the base-band unit which receives/transmits the $B$ bandwidth signals for each of the $M$ antennas, and performs digital computations O($MB$) as MIMO processing to generate the $N$ user streams. 
Hence, mMIMO energy consumption is antenna proportional ($M\times$) to just get user-proportional ($N\times$) data bits across, thus super-linear energy consumption! 
Furthermore, the above mMIMO architectures is not flexible, that is they can not turn down energy consumption proportionately if the number of users drop. 
This is because if mMIMO turns off a certain number of RF chains, it has to drop some antennas, which comes at performance degradation by not using the full array ~\cite{onoff1,onoff2,onoff3,onoff4}.
Thus, traditional mMIMO architecture does not posses a trivial design knob, that can be tweaked to adapt it to differing user-loads without dropping performance, hence lacks flexibility.

In this paper we break the antenna-proportionate energy consumption required to achieve mMIMO multiplexing to user-proportionate energy consumption, while continuing to use all antennas for mMIMO operation, and achieving flexibility (dropping energy consumption as users drop) and spectral efficiency (using only bandwidth B).
To this end, we present the design and implementation of a new massive-MIMO architecture, dubbed as \name.
\name connects each of the $M$ massive-MIMO antennas to an ultra-low power per-antenna configurable analog-network, as shown in Fig \ref{fig:intro_arch} (b). This analog network can intuitively apply a unique spreading signature to signal received from each antenna, spreading the signal in a distinct manner across a wider bandwidth with very minimal power ($<1$mW). The analog-network output from each antennae is combined and passed through a single down-conversion chain, such an approach allows \name to add any number of antenna while only consuming minimal extra energy. The combined signal is then sampled at user-proportionate sampling rate ($NB$), to allows for the digital computations, sampling and compute energy consumption at user-proportionate levels. 

The final goal of \name is to construct the $N$ streams each of bandwidth $B$, from the combined signal sampled at $NB$; the $N$ streams when combined digitally with mMIMO processing should isolate the $N$ user data streams. To ensure that the $M$ antennas diversity gains are captured correctly via the analog-spreading network, \name develops a algorithm to choose proper codes for analog-spreading, such that multiple antennas are grouped strategically to create $N$ virtual RF chains. That is, \name superimposes codes of multiple antennas strategically such that they beamform via the analog network towards one user per virtual RF chain, such that when combined with the MIMO processing it creates the required $N$ user data streams.  
The digital de-spreading and analog-spreading operation can be conducted for transmission as well, to achieve user-proportionate energy consumption for both receive and transmit. 
Furthermore, \name energy consumption can be scaled flexibly via sampling bandwidth design knob, akin to frequency multiplexing.
That is, if the number of users drop to $K$ users, the combined signal across the $M$ antennas are digitized at the $KB$ sampling rate.





To achieve \name architecture, the first challenge is to how to create these spreaded bandwidth signatures in the analog-network which allows creation of the said $N$ virtual RF chains.
Furthermore, we need to achieve this spreading operation in an ultra-low power fashion, which would allow \name to add more antennas easily with minimal power overhead. 
\name's solution to the above problems is to pass each antenna's signals
via  `faster-than-bandwidth RF switches' with active power draw $<1$mW, and combine all the switched outputs into a single unified RF signal.  
The switches modulated the antenna signals with sub-sample level time period binary on-off waveforms, which create the desired analog spreading signature. 
These spreaded signatures can then be configured to control how different antenna groups get combined in the analog network, before the unified switched outputs go through the shared downconversion+digitization interface in the single RF chain.

The second challenge in \name architecture is how does \name  isolate these $N$ virtual RF chains emerging from via the shared downconversion and digitization interface?
The insight of \name here is that since we have $NB$ bandwidth at disposal, and the user signals are $B$ bandwidth, the spreading codes (used in analog domain) can be orthogonalized in time domain. 
This allows the time-domain digital de-spreading block to splice the $NB$ digitized samples at the orthogonalized time samples to isolate the $N$ virtual RF chains from the shared digitized analog network output. 
This time domain de-spreading ensures that the de-spreading operation is optimal and reverts any distortions created in the analog spreading effect. 
This enables the created virtual RF chains to have similar performance as what a physical RF chain would achieve. 
That is, even though there did not exist a physical path to $N$ antennas, it gets created virtually post-digitization via the analog spreading+ digital de-spreading and behaves as a physically laid chain. Furthermore, \name achieves flexibility; we can use design knob of variable sampling bandwidth to vary the number of virtual chains as number of users $N$ varies.

A natural question here is that, does \name split $M$ antennas across $N$ virtual chains?
In other words, because of the antenna grouping, are $M$ antennas shared in a non-overlapping manner across these virtual chains?
In fact, because of the virtualized behaviour, infact the entire $M$ antenna array is available to all the $N$ virtual RF chains.
The two key insights here are that virtual RF chains allow for both, multiple antennas to be connected to a single virtual RF chain, as well as a single antenna connected to multiple RF chain.
This is made possible by the orthogonality of the codes, in the way that we can add to codes together for an antenna so that it shows up in both the spliced samples of virtual RF chain created by both these codes.
Further, giving same spreading code to multiple antennas allows them to be connected together in the same virtual chain.
This way, we can have one virtual RF chain using a group of antennas which are co-phased for a user, and grouping them would basically beamform towards that particular user.
This allows \name to create narrower beams by using more number of $M$ antennas, than $N$ virtual chains, with the many-to-many connection allowing all antennas to be shared across all virtual chains.
Hence, \name achieves mMIMO like performance at user-proportionate power for $N$ virtual RF chains without being antenna-proportionate as mMIMO.

We implement \name hardware and software prototype implementation with $M=8$ antennas to flexibly serve $2,3,4$ users, as desired by network provider. We deploy \name in indoor environments over multiple user locations, and benchmark against the traditional mMIMO architecture and frequency multiplexing(OFDMA).
Our key results demonstrate similar performance metrics as compared with traditional mMIMO architecture with $3\times$ less power requirements, achieving the same network capacity as compared to frequency multiplexing with $4\times$ lesser spectrum requirements, which experimentally confirm the spectral efficiency and user-proportionate energy consumption. 
At the same time, if there is a single MIMO capable user, \name can be used to serve multiple streams, hence increasing the throughput of single user by $3$ times in energy proportionate manner.
Finally, we end with a case study on how \name can save upto $40$\% power in a 5G NR base station.

\ifsigcomm
\else

\section{Can Massive MIMO achieve User-proportionate energy?}

Before delving into the design details of \name, which creates a new user-porportionate MIMO architecture while interfacing many more antennas than users, we will briefly go over background on existing approaches.

The biggest roadblock to achieving user proportionate energy consumption, is that we fundamentally need many antennas than number of users, in order to create the required narrow beams to isolate each users signal using the same spectrum.
This is evident in history of MIMO deployments, which started with mu-MIMO (multi-user MIMO) proposing to use only $N$ antennas for $N$ users and was deployed in LTE and Wi-Fi standards\cite{kuhne2020bringing,castaneda2016overview,liu2012downlink,pasandi2021latte}. mu-MIMO was largely unsuccessful, and got replaced by massive MIMO (mMIMO) in the next generation 5G networks\cite{ericonmmimo,huaweiwhite,ding2020agora}. The primary reason for this is with $N$ antennas the beams created aren't narrow enough to isolate $N$ users.

Thus, mMIMO proposes using massive number of antennas $M$ to serve $N$ users, $M>>N$, to create narrow beams and resolve the issues with mu-MIMO.
To do so, mMIMO has traditionally adopted the full-digital architecture\cite{ding2020agora,yan2019performance,araujo2016massive, hoydis2013massive,liu2012downlink, castaneda2016overview}, which requires digitization of signals received at each antenna, hence each antenna needs its own separate RF chain. This makes the power consumption antenna-proportionate ($M\times$) ~\cite{mondal201825,mondal201725,mondal201921,barati2020energy} just to get $N\times$ spatial multiplexing, thus being energy inefficient. 
To address the very high data communication needs of $M$ antenna signals to the BBU, modern mMIMO deployments do the digital processing at radio front end to get $N$ user-streams and only backhaul those user-streams instead of per-antenna streams \cite{ding2020agora,ericonmmimo}, as seen in Fig. \ref{fig:intro_arch} (a). This makes mMIMO inflexible, since the per-antenna RF chains are not exposed to higher layers it keeps on burning the high antenna-proportionate power at baseband unit and does not adapt to varying user-load. 
This is evident in power consumption trends of 5G base stations with mMIMO, where baseband power consumption dominates power consumed by amplifiers in RF front ends \cite{huaweiwhite}.


An alternate to fully digital architecture is to offload some power to analog domain by creating `Hybrid beamforming (HBF)' architecture to achieve the goal of reducing number of RF chains. 
To do so, HBF's connect $M>>N$ antennas to $N$ RF chains via a compressive analog network (usually consisting of phase shifters) which maps the `useful' aspects of $M$ antennas into the $N$ streams, with a large body of works on creating different analog mappings like fully-connected, partially-connected and choosing the network configurations strategically \cite{kim201828,kuhne2020bringing,ghasempour2018multi,xie2015hekaton,chen2020turbo, zhang2017hybridly,yu2016alternating,mondal201921, mondal201825, mondal201725,obara2015indoor, chen2017bush}.
These analog networks make HBF's rigid, as they are constructed to optimize power consumption for fixed $N$, and can not be adapted to vary the number of RF chains if user load varies.
In addition to rigidity, in practical HBF deployments the analog compressive network has high insertion losses, both due to the hardware inefficiency, as well as fundamentally due to large amounts of signal splitting to create the complicated $M\to N$ networks (can go >10dB \cite{kuhne2020bringing,yan2019performance,barati2020energy}).
These losses need to be compensated appropriately with separate amplifiers \cite{barati2020energy} required per-antenna, thus making the RF front end power consumption antenna-proportionate and hence reducing EE.
With more antennas, the analog network gets more complicated and the insertion losses start increasing, and there are practical studies showing that HBF consume almost similar power as compared to fully digital architectures \cite{yan2019performance,barati2020energy}.

Hence, in contrast to existing classes of fully digital and HBF approaches, \name creates a new class of beamformers where an ultra-low power configurable analog network basically helps with virtualization of RF chains in digital domain.
This allows \name to adaptively do $M\to N$, or $M\to N'$ compressive analog-digital mapping, depending on user load, to achieve user-proportionate power while using more antennas.
\name achieves this architecture by interfacing this configurable analog network to a single RF chain, whose sampling bandwidth decides how many virtual RF chains are created.
This greatly simplifies the analog network drastically and reduces insertion losses, and allows
\name to carve $N$ `virtual' RF chains from this $1$ single physical chain, with capability of changing $N$ simply via software, in response to varying user loads.
This way, \name truly achieves the user-proportionate power consumption in a flexible manner.


\section{Design}
\begin{figure}[t]
    \centering
    \includegraphics[width=0.5\textwidth]{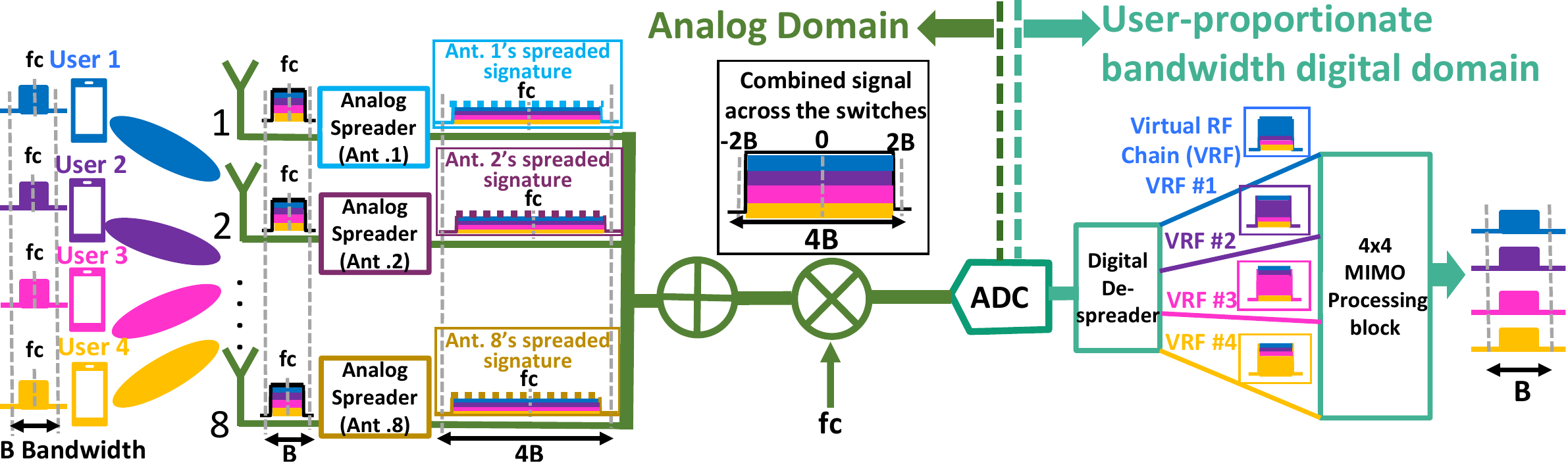}
    \caption{Design overview: \name's analog spreading, digital de-spreading creates the required $N$ virtual RF chains prioritizing $1$ user each, and then can be fed to standard MIMO processing block to procure the $N$ per-user signals}
    \label{fig:intro_spread}
\end{figure}
\name designs a new multi-user MIMO antenna array architecture which maximises both energy and spectrum efficiency (EE \& SE) by adapting to varying user loads and carving out user-proportionate number of virtual RF chains as demand-fit from a single physically laid RF chain.
The key design elements which enable \name are the analog spreading, digital de-spreading blocks which allow creation of these virtual RF chains, and BABF approach to beamform towards one user-per virtual chain via many more antennas than number of users (Design overview visually illustrated in Fig. \ref{fig:intro_spread}).
In this section, we will first describe how \name realizes the analog spreading via RF switches, how it is de-spreaded in digital domain, then describe the BABF approach. 
Finally, we end with how all these design elements work together to handle the multi-user interference and enable spatial multiplexing

\subsection{How \name realizes analog spreading via RF switches}

To practically realize the analog spreading block as envisioned in Fig. \ref{fig:intro_spread}, \name needs to consider two requirements.
First requirement, is that the spreading effect has to be created right next to antenna in the RF domain, before the signals across the antennas have been combined.
Hence, it rules out any simple baseband freq. shift circuits which implement frequency spreading.
Second requirement is that this circuit element should have almost-zero, or minimal power consumption (denoted as $P_{\text{analog spreading}}$) as compared to power in the single OFDMA like RF chain (denoted as $P_{\text{OFDMA}}$).
This is because $P_{\text{GreenMO}}=MP_{\text{analog spreading}}+P_{\text{OFDMA}}$, as $M$ antennas are passive and only other active elements apart from the OFDMA RF chain are these switches.
Thus, we need to have $P_{\text{analog spreading}}<<P_{\text{OFDMA}}$, so that $P_{\text{GreenMO}} \approx P_{\text{OFDMA}}$, and \name achieves the same EE as OFDMA.

A naive approach here would be to use frequency mixers to create the desired frequency spreading effect. 
The mixers can have the LO clock sources of the order of signal bandwidth, instead of center frequency, to shift the signals across the $NB$ spectrum.
Hence, as compared to mixers typically used for downconversion operation, these $NB$ mixers would be shifting frequencies in order of bandwidth ($O(B)$) instead of center frequency ($O(f_c)$), and hence will have orders of magnitude lower active power draw, which is proportional to frequency of operation. This would be because the $B$ is just a small fraction of $f_c$, for eg, in Wi-Fi  $f_c= 2.4/5$ GHz whereas $B\sim 20$ MHz.
Although mixers by themselves can be passive circuits, they typically have considerable frequency conversion losses of about 5-10 dB\cite{mixer1,mixer2}.
Thus, a mixer spreading unit would require per-antenna amplifiers (PA/LNA) to offset the losses. 
Hence, even though mixers can work right at the RF level, it does not fit the second requirement due to the conversion losses.

Instead, \name's insight is that both the requirements can be met by taking a leaf out of backscatter systems, which face similar constraints on RF circuits which spread frequencies to nearby channels \cite{zhang2016hitchhike,zhang2017freerider,dunna2021syncscatter,liu2020vmscatter,kellogg2014wi}.
These backscatter systems also need near zero power power operation to justify batteryless operation, and can not tolerate insertion losses which would lead to reduced reflected signal power.
As a consequence, backscatter systems use simple RF switches, instead of mixers to meet these requirements.
Clearly, RF switches work directly at the RF frequencies as they have the capability to toggle the impinging RF signals on and off.
Further, this on-off toggling if done periodically with a on-off square wave of certain frequency create harmonics and spreads the signals.
Also, RF switches have minimal insertion losses, <1 dB, and hence do not require amplifiers and as an evidence have been used successfully in backscatter systems without needing LNAs/PAs.
\begin{figure}[t]
    \centering
    \includegraphics[width=0.5\textwidth]{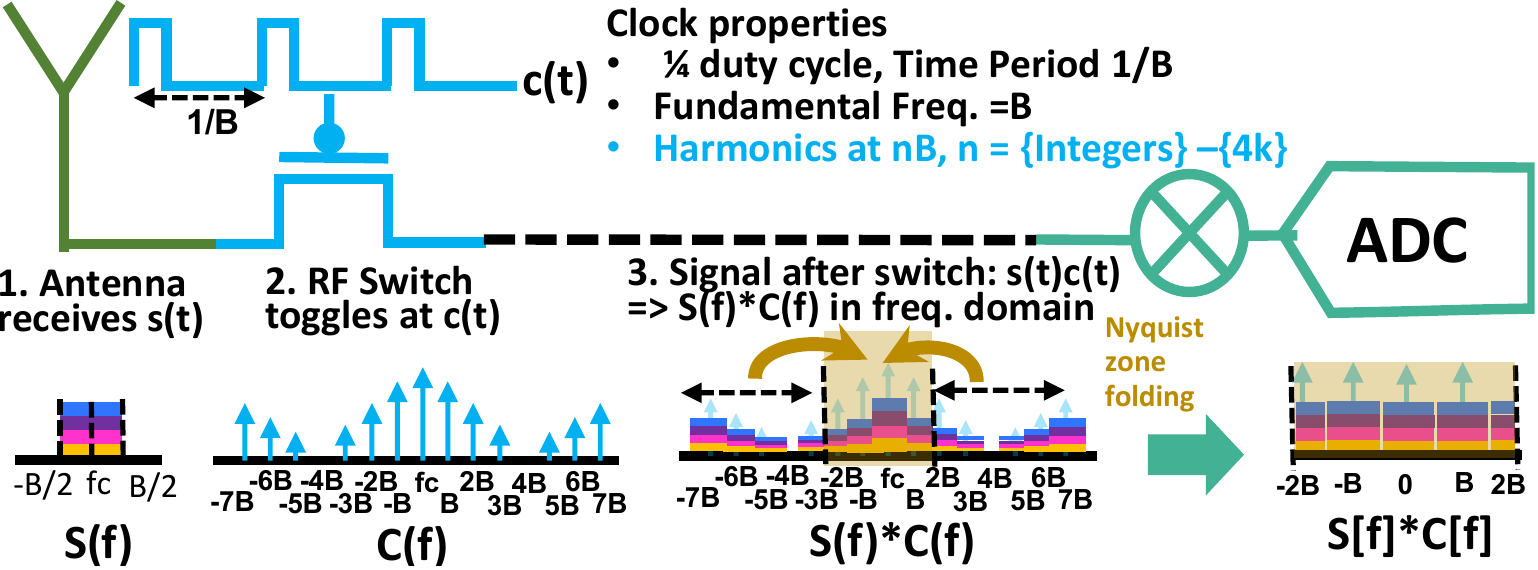}
    \caption{How RF switches create the required signal spreading effect}
    \label{fig:spread_des_figa}
\end{figure}
To explicitly show how \name uses RF switches as analog spreading unit, we model the signal at antenna to be $s(t)$, which goes through RF switch toggling a periodic on-off wave given by $c(t)$. 
Due to switching, we get multiplication in time domain $s(t)c(t)$, corresponding to convolution in frequency domain $S(f)*C(f)$.
If we take a $1/B$ time period on off sequence $c(t)$ it will fundamental frequency as $B$. 
Thus, the active power draw in RF switch would be $O(B)$ instead of $O(f_c)$, since the time period is $1/B$.
As a consequence, the RF switch would work at orders of magnitude lower power than downconverters which require LO clocks with time period $1/f_c$.
Thus, the RF switches are able to work at RF level and with minimal power requirements, both from the insertion losses, as well as active power draw.
However, the $B$ frequency on-off codes will have harmonics at integral multiples of $B$ which would spread the signal much beyond the nyquist period $[-NB/2,NB/2)$.
This is visually illustrated in Fig. \ref{fig:spread_des_figa} for $N=4$.

So a natural question is how do we obtain the required spreading effect only between $[-NB/2,NB/2)$ and remove the non-linearities?
A naive solution of to basically do a low pass filter for the band of interest $[-2B,2B)$ before sampling to eliminate these copies altogether.
However, this would end up wasting the signal power which has landed beyond the band, as these wold just get filtered out.
Instead, we realize that the non-linearities created by switching can infact be harnessed in a powerful way by simple sampling process.
We make the observation that by sampling at $NB$ rate, a $1/N$ duty cycled code of $B$ frequency basically gives equal harmonics at integral frequencies in the required nyquist range.
For more details please refer to mathematical proof \cite{proofs}.
This is because the other harmonic components simply alias on top of the required peaks in the $[-NB/2,NB/2)$ (shown visually via folding arrows in Fig. \ref{fig:spread_des_figa} for $N=4$).
Hence, instead of filtering these non-linearities, by simply sampling them via $NB$ ADC these harmonics fold on top of each other and thus make the system efficient by not wasting any received signal power in filtering.

Hence, while serving $N$ users, using RF switches, we can make use of the required $NB$ bandwidth to handle these $N$ users by spreading each antenna signals into this wide bandwidth via RF switches, at minimal power overhead and efficiently without losing any signal power.
This is possible by careful duty cycling of the switching clock, such that the created harmonic distortions alias on top of each other to make the spreading process efficient.

\subsection{\name's digital de-spreader to create $N$ virtual RF chains}
\label{sec:virt_rfc}

So far, we have seen how the RF switches can act as minimal-power analog spreading units.
Now, we will show how the digital de-spreader works to isolate the $N$ individual per-antenna signals, which in abstract notion can be thought of as creating $N$ virtual RF chains towards each antenna.
Even though there does not exist a physical RF chain to each antenna, the isolation of per-antenna signals in digital domain from the net $NB$ bandwidth would lead to creation of virtual chains interfacing each antenna to the ADC, without any actual physical laid hardware.
In this subsection we will simplify the setting to one antenna per virtual chain (So $M=N$), and the next subsection generalizes to multiple antennas per virtual chain, so that $M>N$.

The basic idea behind digital de-spreading is that the $NB$ bandwidth allows creation of $N$ time-orthogonal spreading codes, which can then be de-spreaded as these codes do not overlap with each other.
Each of these orthogonal spreading codes have $1/N$ duty cycle and same frequency $B$, however with different initial phases (As shown in continuous time code waveforms, Fig. \ref{fig:duty_fig}).
As a consequence, when these codes are sampled with $NB$ bandwidth, these codes basically represent different sample indexes in time domain, since each time sample occupies $1/NB$ time, and because of $1/N$ duty cycling basically each spreading code turns `on' an antenna for different on-off samples (As shown in discrete time sampled code waveforms, Fig. \ref{fig:duty_fig}).

Essentially, $c_0$ is on for every $Ni$ samples, $c_1$ is on for $Ni+1$ samples and generalizing $c_j$ is on for $Ni+j$ samples.
Hence, the digital de-spreader needs to isolate these samples to `invert' the codes, since by collecting every $Ni+j$ samples it removes the `off' samples of $c_j$ and preserves only the `on' samples, and hence, removes all the harmonics and preserves only the modulated content.
However, this requires sampling synchronization with respect to a particular $c_i$, which we choose to be $c_0$ fairly generally.
In our implementation, we ensure this by deriving $c_0$, and other clocks from the SDR's sampling frequency.
This is illustrated visually via Fig. \ref{fig:des_figc}.

We can similarly de-spread by inverting the codes in frequency domain.
It is because, when the fourier transform of these discrete sequences are taken, though the magnitude spectrum is identical, i.e. all the $N$ codes show $N$ delta functions at every integral multiple of $B$ between $[-NB/2,NB/2)$, the phase response is distinct because of different initial phases.
Each $c_i$ shows a different phase at these delta functions, with the phases following a routes of unity sequence, like $c_i$ has phases $\frac{\pi}{N}*i*j$ for different $j$ denoting the $N$ delta functions $-NB/2,-(N-1)B/2,\ldots (N-1)B/2$.
Please refer to \cite{proofs} for detailed mathematical proofs.
The codes and their frequency components are shown visually in Fig. \ref{fig:duty_fig}.
However, as explained before, the same de-spreading operation is achieved more intuitively in the time domain.
\begin{figure}[t]
    \centering
    \includegraphics[width=0.4\textwidth]{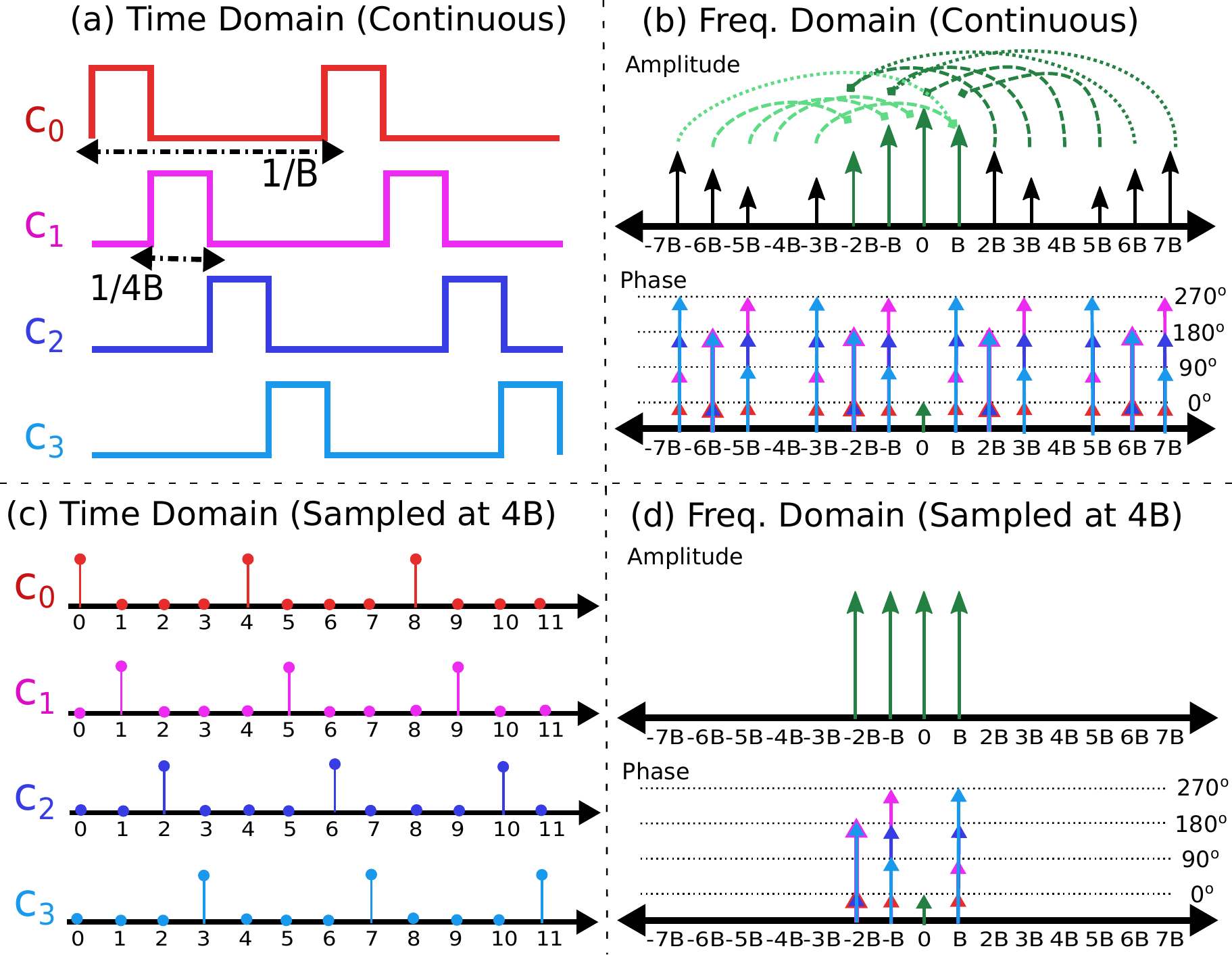}
    \caption{\name on-off codes: continuous and discrete time }
    \label{fig:duty_fig}
\end{figure}

\begin{figure}[t]
    \centering
    \includegraphics[width=0.5\textwidth]{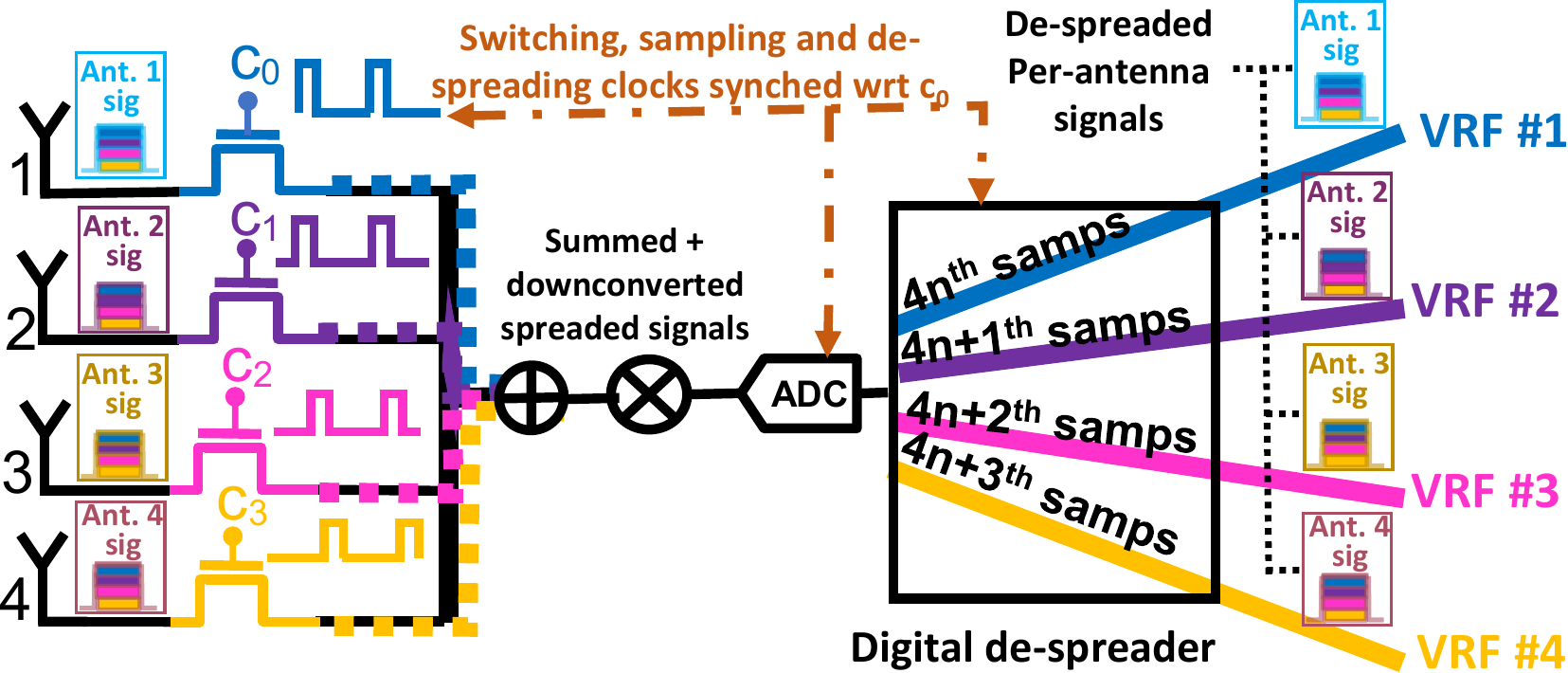}
    \caption{How \name's digital de-spreader works}
    \label{fig:des_figc}
\end{figure}


Hence, as illustrated in Fig. \ref{fig:des_figc}, even after the signals from multiple antenna get combined and pass through a single downconversion chain, by just isolating every $Nk+j$ samples, we can get the samples corresponding to $j$-th antenna. 
Thus, in a way this switching+sampling operation enables separate signal path for each antenna, in a way the downconversion chain is virtualized over all the antennas.
This makes \name architecture with virtual RF chains a flexible architecture, as \name can adjust the number of virtual RF chains on the fly. 
That is, if the number of users change from $N \to N'$, the architecture can respond by simply changing the sampling rate from $NB$ to $N'B$, which can then be used to carve out the needed $N'$ virtual RF chains.
In comparison, a traditional MIMO architecture can not increase/decrease number of RF chain in response to user demand as these RF chains have been laid physically, and can not be changed on the fly.
In a way this $M=N$ antenna version of \name enables a user-demand `flexible' version of standard digital beamformer used traditionally in Massive MIMO, as it isolates one antenna signal per virtual chain.

However, \name goes beyond just mimicking a digital beamfomer, and can enable green communications with user-proportionate power consumption.
To justify this, \name needs to create 'user-proportionate' number of virtual chains, instead of requiring number of virtual chains same as number of antennas.
Hence, we generalize the architecture for varying $N$ users with $M>N$ antennas in the next subsection, so that the sampled bandwidth remains $NB$ while increasing antennas to $M$. 
With this generalization, we truly show how \name architecture can flexibly meet the user-proportionate power consumption demands.

\begin{figure}[t]
    \centering
    \includegraphics[width=0.5\textwidth]{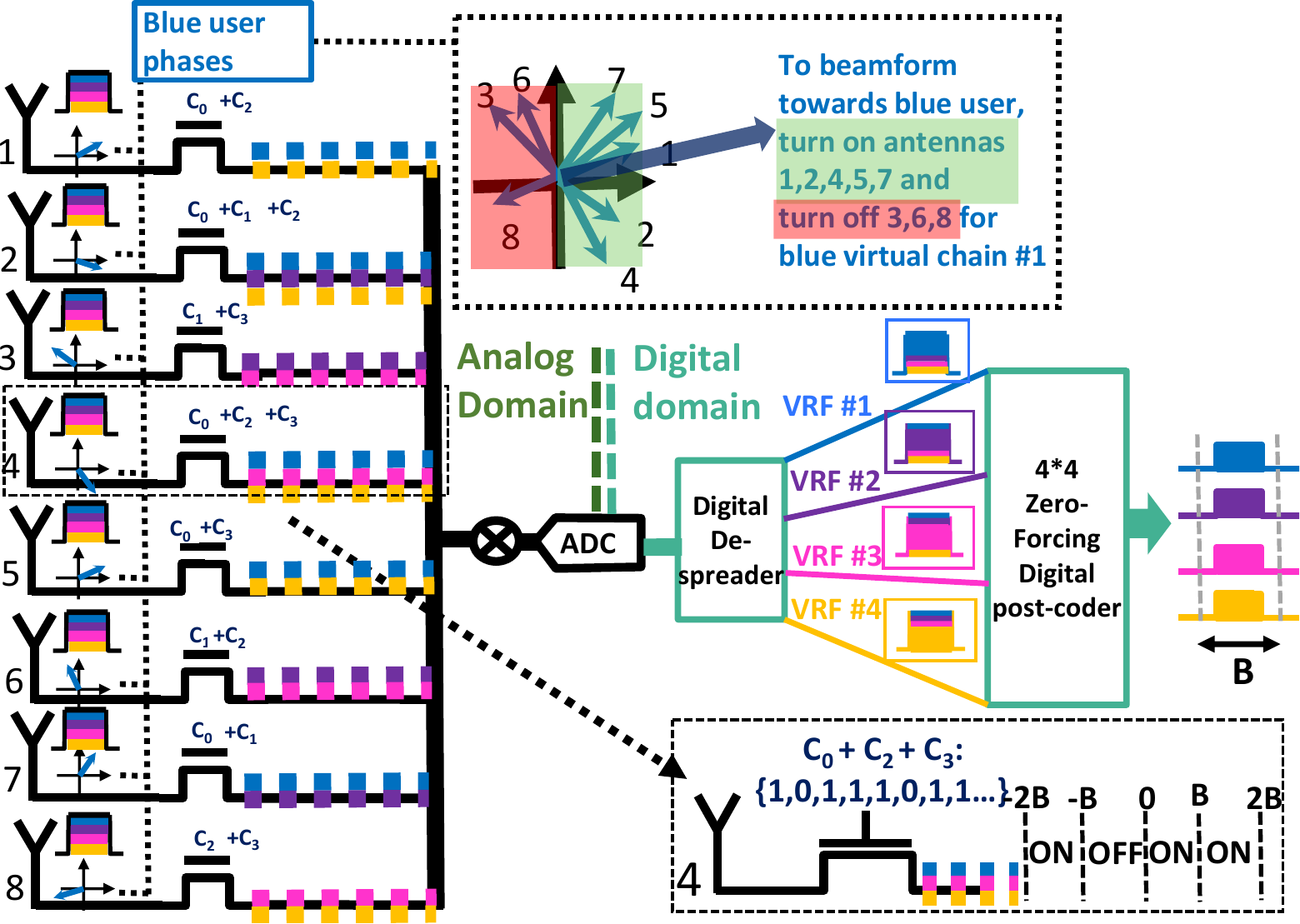}
    \caption{Generalizing \name to multi-antenna architecture}
    \label{fig:des_figd}
\end{figure}

\subsection{How \name flexibly generalizes to $M$ antennas for $N$ users}

Having described how analog spreading, digital de-spreaing allows \name to create $N$ virtual RF chains, we will now show how we can interface many more antennas $M>N$ to these $N$ virtual RF chains.
The key insight to this generalization is the fact that unlike a physical RF chain, which is connected to one antenna typically, the virtual RF chains are by default connected to all the RF chains simultaneously.
In other words, we can turn `on' multiple antennas per virtual RF chain instead of a single antenna, and also, an antenna can simultanoeusly be `on' for multiple virtual RF chains.
This multiple antenna to multiple virtual RF chain mapping is enabled by the orthogonality of the toggling sequences $c_j$.
That is, if we want to turn on antennas $i_1,i_2,i_3$ for the virtual RF chain created by $c_j$, we can supply the $c_j$ clock to each of these antennas indexed via $i_1,i_2,i_3$.
And if a particular $i$-th antenna has to be turned on for $j_1,j_2,j_3$ virtual RF chains, we can supply clocks $c_{j_1}+c_{j_2}+c_{j_3}$.
Since $c_{j_1}$, $c_{j_2}$ and $c_{j_3}$ don't overlap in time, adding these codes together would create a new toggling sequence which would be `on' for $Nk+j_1,Nk+j_2,Nk+j_3$ samples and hence turn on this antenna for all these $3$ virtual RF chains.
This is illustrated visually via right-bottom inset in Fig. \ref{fig:des_figd}.

\begin{figure}[t!]
    \centering
    \includegraphics[width=0.5\textwidth]{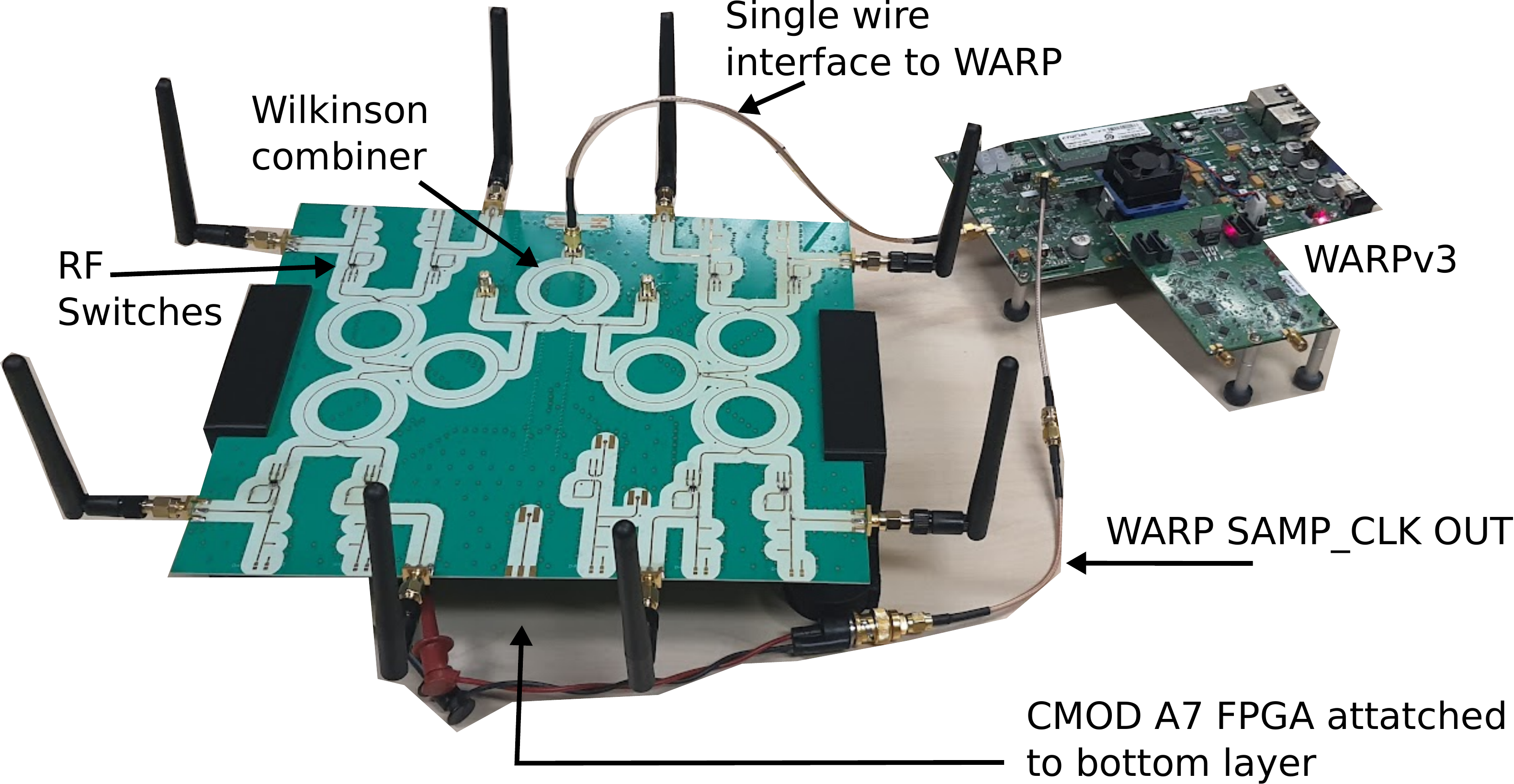}
    \vspace{1pt}
    \caption{Hardware implementation of \name}
    \label{fig:pcb_pic}
\end{figure}

Putting this in a matrix form, because of this many to many mapping, where multiple antennas can be turned `on', `off' for multiple virtual RF chains, the switching network allows us to implement a binary matrix $\mathbf{S} \in \{0,1\}^{M*N}$ in analog domain, where $M$ is number of antennas and $N$ is number of users, as well as number of virtual RF chains to justify the user-proportionate computing claim.
Basically, this matrix projects the higher dimensional channel $\mathbf{H}$ which is complex valued $\mathbf{C}^{N*M}$ which represents the amplitude and phase of each $n$
-th user at $m$-th antenna, into an equivalent $\mathbf{C}^{N*N}$ chnanel $\tilde{\mathbf{H}}=\mathbf{HS}$ by toggling antennas on-off strategically.

Hence, \name architecture can project the $N*M$ over-the air wireless channel between $N$ users and $M$ antennas into $N*N$ equivalent channel between the $N$ users and $N$ virtual RF chains, via $\mathbf{S}$ matrix implemented in analog domain by RF switches and the supplied clocks.
The choice of this matrix when $M=N$ is obvious, we can just set it to identity matrix and continue, as was also motivated in the prior sections.
When $M>N$, we need to select the matrix $\mathbf{S}$ strategically such that we tap into the higher degrees of freedom offered by more antennas and project the higher dimensional matrix $\mathbf{H}$ into a well-formed $N*N$ equivalent $\tilde{\mathbf{H}}$.
To do so, our insight is that since we have $N$ virtual RF chains for $N$ users, we can user the high number of antennas to perform approximate on-off beamforming towards one user per virtual RF chain.
That is, for virtual RF chain $j$, we basically turn `on' the maximal set of antennas in-phase for user $j$, so that the signal power for user $j$ is boosted for $j$-th virtual chain.
This is visually illustrated via right-top inset in Fig. \ref{fig:des_figd}, with an example of virtual RF chain antenna configuration for blue user.
We dub this selection of the matrix $\mathbf{S}$ which implements per-user beamforming as BABF approach (Binarized analog beamforming).
The BABF approach basically takes the analog beamforming weights and quantizes it to binary $0-1$ level, so that it can be implemented in analog domain via RF switches.

\textbf{Putting it all together}: So, to conclude the design section, we will briefly summarize and put the various design elements in context.
We showed how the analog spreading, digital de-spreading can efficiently use the $NB$ bandwidth of single RF chain to enable $N$ virtual RF chain.
Then we showed how the per-antenna codes can be chosen such that the entire $M$ antenna array is available to all the $N$ virtual RF chain, by means of the switching matrix $\mathbf{S}$.
This allows \name to use the array to beamform towards one user per virtual chain, via the BABF approach. 
This analog spreading + digital de-spreading + BABF approach culminated in creation of an equivalent $N*N$ channel $\tilde{\mathbf{H}}=\mathbf{H}\mathbf{S}$ between the $N$ users and $N$ virtual RF chains.
This $\tilde{\mathbf{H}}$ can then be fed to digital beamformer block which would invert this $\tilde{\mathbf{H}}^{-1}$to finally get the per-user signals.

\begin{figure*}[t]
    \centering
    \includegraphics[width=0.9\linewidth]{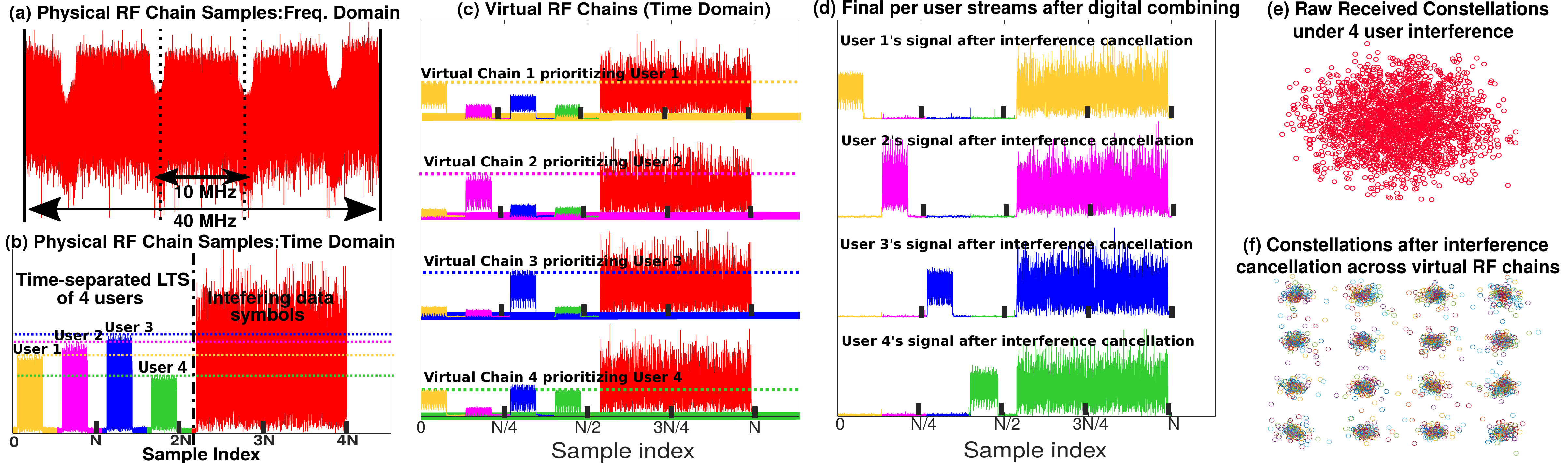}
    \caption{(a) Shows the raw received signal from the single RF chain. The created virtual RF chains are shown in (b), after BABF which increases user $i$'s power in $i$-th virtual chain. Finally (c) shows the interference free per-user streams obtained after digital combining of the virtual chains. (d) and (e) show the raw constellations before and after interference cancellation.}
    \label{fig:ex_trace}
\end{figure*}

\fi


\section{Implementation}\label{sec:implementation}


We implement the \name architecture on a custom multi-layer PCB prototype fabricated using Rogers substrate (as shown in Fig. \ref{fig:pcb_pic}), with HMC197BE\cite{HMC197BE} RF switches for analog spreading and CMOD A7 15t FPGA \cite{cmod} to generate the on-off $1/N$ duty-cycled clocks $c_i$ for digital de-spreading.
On the top-layer of the PCB, we have the RF plane of \name architecture, consisting of interfaces to utilize COTS antennas via SMA connectors, RF switches, wilkinson combiner networks to combine the 8 switched antennas into a single wire interface.
On the bottom layer of the PCB, we have the control plane of \name architecture, consisting of CMOD-A7 FPGA attached to the PCB via header pins, which provides the $c_i$ codes to RF switches on top layer using vias. 

We utilize WARPv3 as the SDR for our implementation of an uplink receiver. The sampling clock for WARPv3 is $40$ MHz, thus a sampling time period of 25 ns.
The HMC197BE RF switch has tRISE = 3 ns, and tON = 10 ns, sufficiently lesser than the sampling time period of 25 ns, hence the selected switches are fast enough to perform the bandwidth level frequency switching for analog spreading. 
In addition, the CMOD and WARPv3 sampling clocks are synched to ensure digital de-spreading works as shown in Fig. \ref{fig:des_figc}.
We achieve this synched behaviour by writing custom verilog modules on the CMOD FPGA which derive the switching clocks $c_i$ from WARPv3's sampling clock, which is interfaced to the FPGA via WARP's CM-MMCX module jumper pins.

In our hardware implementation, we test for $M=2,3,4$ users.
Since the sampling bandwidth of WARPv3 platform is $40$ MHz, the per-user bandwidth is fixed to $10$ MHz to be able to support $4$ users.
These $4$ 10 MHz users are implemented via $4$ independent USRP SBX daughterboards which do not synch their clocks with WARP receiver.
Hence, at a maximum, we need to create $M=4$ virtual RF chains, we would need to switch different antennas on-off for every $\{4n\}$-th, $\{4n+1\}$-th, $\{4n+2\}$-th and $\{4n+3\}$-th samples. 

Since our PCB implementation has $8$ antennas, the switching matrix $\mathbf{S}$ is a $8\times M$ matrix, with each $i$-th row representing the on-off states of the $i$-th antenna for the $M$ different virtual RF chains. 
For an example, say $M=4$ and this row was $[1,0,1,0]$, so this would simply be implemented as $1*c_0+0*c_1+1*c_2+0*c_3$.
Thus, in order to implement this matrix in hardware, we represent each row, which is a $M\times 1$ binary vector by a hexadecimal digit, and communicate $8$ of these hexadecimal digits representing antennas' on-off states to the FPGA via a standard UART code. 

\subsection{\name's example captured over-the-air trace}


To implement the required digital signal processing, we utilize the WARPLab codes in MATLAB to utilize 802.11 compliant OFDM waveform with 64 subcarriers (48 data, 4 pilots and 12 null subcarriers). 
An example received trace collected via WARPv3 from 4 interferers is shown in Fig. \ref{fig:ex_trace}.
By plotting the frequency domain spectrum of the sampled signal via WARP, we can see the anlog spreading in action which has taken the $10$ MHz user signals and spread it to $40$ MHz (Fig. \ref{fig:ex_trace} (a)).
When plotted in time domain, first we can see via different colors in Fig. \ref{fig:ex_trace} (b) the MIMO preambles with each user having non-overlapping LTS's to allow for channel estimation. However, all the users overlap transmissions of actual data bits, as shown in red.
Note that the single received trace has sample index from $0$ to $4N$ since the received signal bandwidth is $40$ MHz so for $N$ actual samples the receiver get $4$ times that.

From this spreaded sampled signal, we separate the virtual RF chains via time-domain digital de-spreading as discussed in Section \ref{sec:virt_rfc} by isolating $\{4n\},\{4n+1\},\{4n+2\},\{4n+3\}$ samples, as shown in Fig. \ref{fig:ex_trace}(c).
Note that each virtual RF chain has sample indexes from $0$ to $N$ since we capture every $4$-th sample but with different starting sample in the process of virtual RF chain isolation.
Due to implementation of switching matrix $\mathbf{S}$ in analog domain, we see that in a virtual RF chain, one user power is prioritized because of BABF switching, by observing the level of each users' LTS. Even the weakest user (green) gets increased power and becomes comparable top other users in its' virtual RF chain. 

Finally, this equivalent $\tilde{\mathbf{H}}=\mathbf{HS}$ channel thus created in the $4$ virtual RF chains is inverted to get Fig. \ref{fig:ex_trace} (d) which clearly shows that the interference is handled since the LTS's of other users are cancelled out.
When the data is decoded after channel inversion, we recover the transmitted QAM-16 constellation from the $4$ users' interfered signals as shown in Fig. \ref{fig:ex_trace} (e), (f). Hence, this example trace shows how \name works end-to-end, it first samples the spreaded $40$ MHz signals, then \name isolates the virtual RF chains by de-spreading.
Due to choice of switching matrix $\mathbf{S}$, the virtual RF chains are configured in a way that using multiple antennas approximate on-off beamforming is performed for one user, and finally the equivalent channel is inverted to get back the per-user signals.

\section{Evaluations}

Having described the design, implementation and showed an example captured trace of \name, we will now go over various experiments performed to verify the design choices and showcase the power + spectrum efficiency of \name. First, we will go over the experimental setting, then show microbenchmarks and ablation studies, \name and end by showing the end-to-end evaluations and brief case studies on power savings entailed via \name architecture.

\begin{figure}[t]
\begin{subfigure}[t]{0.23\textwidth}
        \centering
        \includegraphics[width=\linewidth]{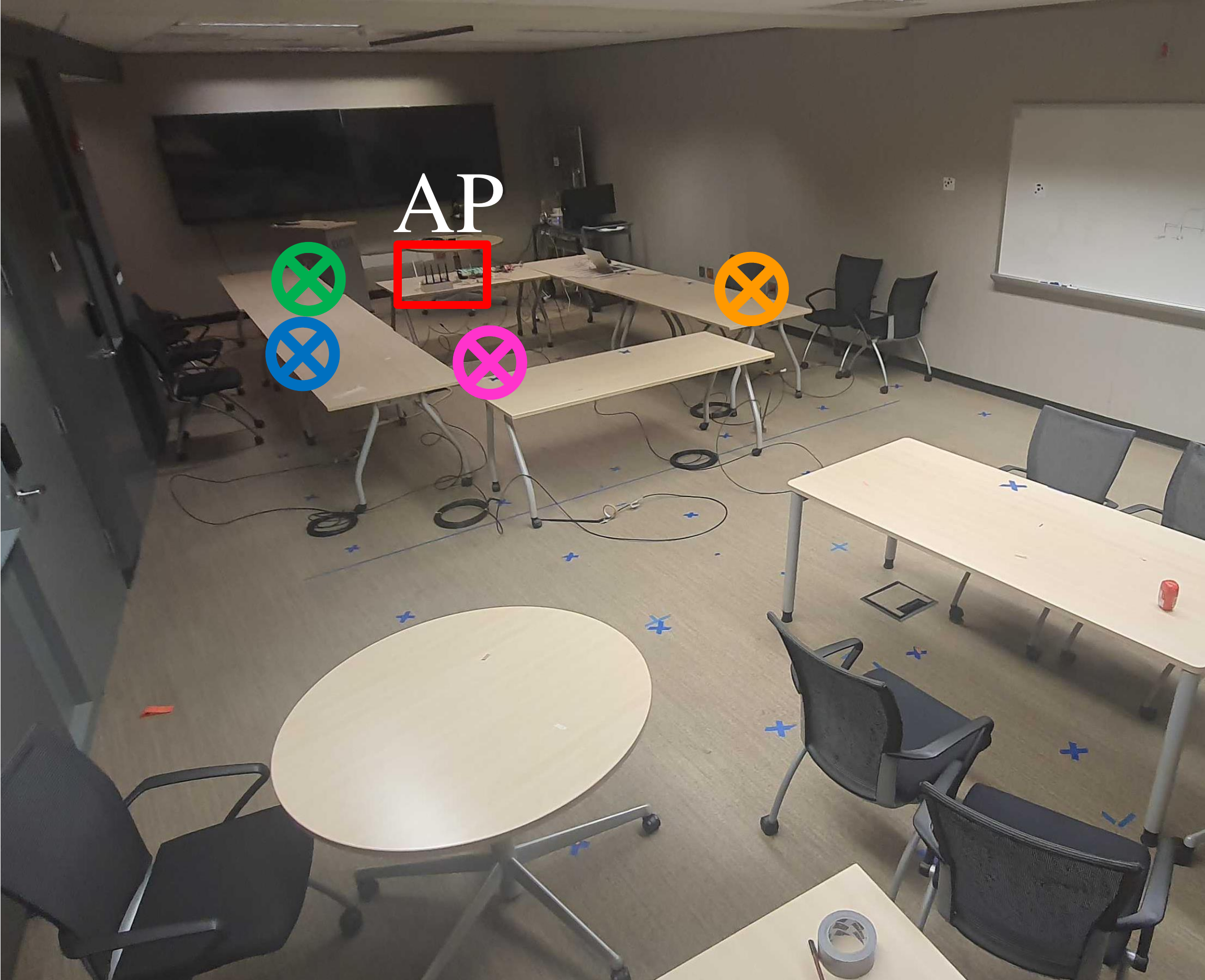}
    \subcaption{Experimental Setting}
        \label{fig:multi_user_conf_actual}
\end{subfigure}
\begin{subfigure}[t]{0.23\textwidth}
        \centering
        \includegraphics[width=\linewidth]{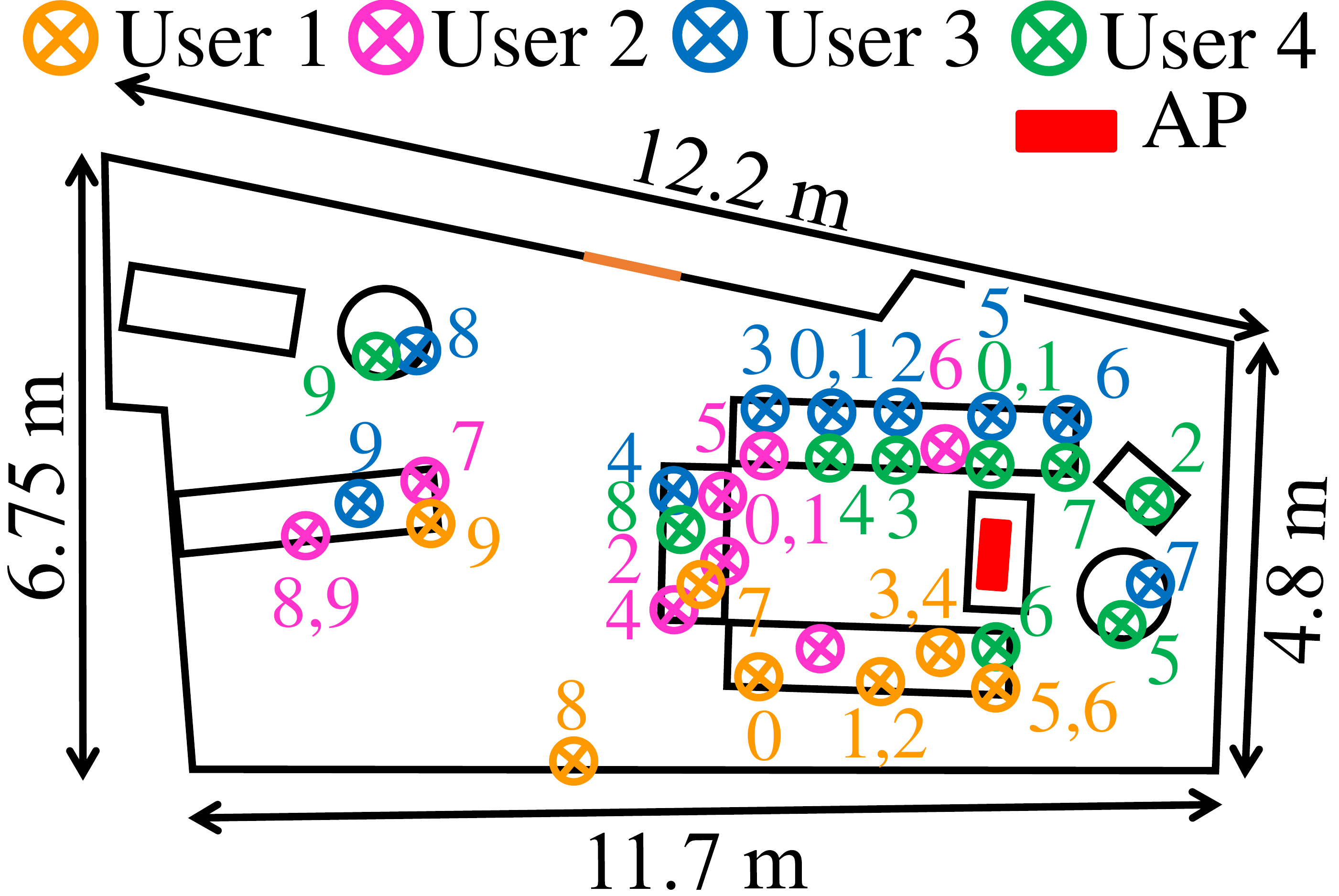}
        \subcaption{Setting Floorplan}
        \label{fig:multi_user_conf_floor}
    \end{subfigure}
\caption{We test \name in a conference room setting (a), and the numbers $0,1,\ldots 9$ in (b) represent the $4$ user positions for the 10 configurations where we test \name and baselines
}
\end{figure}

\textbf{(a) Evaluation setting:} To evaluate \name, we consider a office environment (conference room setting, rough dimensions of 12m*5m, Fig. \ref{fig:multi_user_conf_actual}) with TV screens/desks/whiteboards which would act as reflectors and make it a multipath setting.
We fix the location of \name PCB, which acts as an AP, roughly in middle of the room, and vary the positions of the $2,3,4$ users across 10 configurations scattered around the room, and with various degrees of closeness to capture the overall performance trends in the experimental setting considered.
For the experiments we set the transmit power such that we have average SNR of about 15 dB, and utilize QAM-16 constellation with 0.5 rate convolutional channel code.
Since the average SNR for users' reception is $15$ dB, the net capacity achievable would be $40\log_2(10^{1.5}) \approx 200$ Mbps for interference-free OFDMA approach having $4$ users with $10$ MHz bandwidth each, hence $40$ MHz total.
However, at $15$ dB SNR, the recommended MCS would be QAM-16 with 0.5 rate code\cite{MCS}, which achieves about 48 Mbps goodput for 40 MHz net bandwidth in our implementation, or $4$ 10 MHz users spatially multiplexed.
This choice of SNRs, as well as constellation is consistent with the recent works on Massive MIMO systems \cite{ding2020agora}.
We compare \name with OFDMA, $4$ antenna, and $8$ antenna MIMO baselines.
OFDMA and $4$ antenna MIMO baselines are implemented on WARP as well, however for $8$ antenna MIMO baselines the results are evaluated via trace level simulations by collecting channels from first 4 antennas and then next 4 since WARP does not have interface to support simultanoeus $8$ antenna measurements.

\begin{figure}[t]
    \begin{minipage}[b][][b]{0.2\textwidth}
    \centering
    \begin{subfigure}[b]{\textwidth}
    \centering
    \centering
    \setlength{\figureheight}{0.55\linewidth}
    \setlength{\figurewidth}{0.8\linewidth}
%
%
\begin{tikzpicture}

\begin{axis}[%
width=0.951\figurewidth,
height=\figureheight,
at={(0\figurewidth,0\figureheight)},
scale only axis,
xmin=1.5,
xmax=4.5,
xlabel style={font=\color{white!15!black}},
xlabel={\textbf{Number of users}},
ymin=0,
ymax=6,
ylabel style={font=\color{white!15!black}},
ylabel={\textbf{SE (bps per Hz)}},
axis background/.style={fill=white},
xmajorgrids,
ymajorgrids,
yminorgrids,
legend style={at={(0.01,0.646)}, anchor=south west, legend cell align=left, align=left, draw=white!15!black,
, font=\scriptsize, nodes={scale=0.75, transform shape},
inner sep=0pt},
xlabel style={font=\scriptsize},ylabel style={font=\scriptsize},ticklabel style={font=\scriptsize},
ylabel shift={-6pt},
xlabel shift={-6pt}
]




\addplot [color=green, line width=2pt]
 plot [error bars/.cd, y dir=both, y explicit, error bar style={line width=2.0pt}, error mark options={line width=2.0pt, mark size=2.0pt, rotate=90}]
 table[row sep=crcr, y error plus index=2, y error minus index=3]{%
2   2.31        0.02         0.02\\
3   3.45        0.07         0.07\\
4	4.7100568	0.131312	0.131312\\
};
\addlegendentry{\textbf{\name}}

\addplot [color=black, dotted, line width=2.0pt]
  table[row sep=crcr]{%
2	2.396\\
3	3.594\\
4	4.792\\
};
\addlegendentry{\textbf{Oracle: 8 MIMO}}

\addplot [color=yellow, line width=2pt, mark size=2.0pt, mark=x, mark options={solid, yellow}]
 plot [error bars/.cd, y dir=both, y explicit, error bar style={line width=2.0pt}, error mark options={line width=2.0pt, mark size=2.0pt, rotate=90}]
 table[row sep=crcr, y error plus index=2, y error minus index=3]{%
2	2.29199382022472	0.040786224093029	0.040786224093029\\
3	3.31248881578947	0.228358549465577	0.228358549465577\\
4	3.95620074626866	0.43363140832695	0.43363140832695\\
};
\addlegendentry{\textbf{4 MIMO}}

\addplot [color=red, line width=2pt, mark size=2.0pt, mark=x, mark options={solid, red}]
 plot [error bars/.cd, y dir=both, y explicit, error bar style={line width=2.0pt}, error mark options={line width=2.0pt, mark size=2.0pt, rotate=90}]
 table[row sep=crcr, y error plus index=2, y error minus index=3]{%
2	1.155   0.01 0.01\\
3	1.15   0.03 0.03\\
4	1.12   0.07 0.07\\
};
\addlegendentry{\textbf{OFDMA}}

\end{axis}
\end{tikzpicture}%
    \end{subfigure}
    
    \begin{subfigure}[b]{\textwidth}
    \centering
    \centering
    \setlength{\figureheight}{0.55\linewidth}
    \setlength{\figurewidth}{0.8\linewidth}
%
%
\begin{tikzpicture}

\begin{axis}[%
width=0.951\figurewidth,
height=\figureheight,
at={(0\figurewidth,0\figureheight)},
scale only axis,
xmin=1.5,
xmax=4.5,
xlabel style={font=\color{white!15!black}},
xlabel={\textbf{Number of users}},
ymin=0,
ymax=100,
ylabel style={font=\color{white!15!black}},
ylabel={\textbf{PE (Mb per joule)}},
axis background/.style={fill=white},
xmajorgrids,
ymajorgrids,
legend style={at={(0.01,0.616)}, anchor=south west, legend cell align=left, align=left, draw=white!15!black,
, font=\scriptsize, nodes={scale=0.8, transform shape},
inner sep=0pt},
xlabel style={font=\scriptsize},ylabel style={font=\scriptsize},ticklabel style={font=\scriptsize},
ylabel shift={-6pt},
xlabel shift={-6pt}
]

\addplot [color=green, line width=2pt, mark size=1.0pt, mark=o, mark options={fill opacity=0}]
 plot [error bars/.cd, y dir=both, y explicit, error bar style={line width=1.0pt}, error mark options={line width=1pt, mark size=3.0pt, rotate=90}]
 table[row sep=crcr, y error plus index=2, y error minus index=3]{%
2   42.77   0.13   0.13\\
3   53.62   1.2    1.2\\
4	61.81	2.17	2.17\\
};
\addlegendentry{\textbf{\name}}

\addplot [color=black, dotted, line width=2.0pt]
  table[row sep=crcr]{%
2	43.12\\
3   54.95\\
4	63.55\\
};
\addlegendentry{\textbf{Oracle:} \textbf{OFDMA}}

\addplot [color=yellow, line width=2pt, mark size=2.0pt, mark=x, mark options={solid, yellow}]
 plot [error bars/.cd, y dir=both, y explicit, error bar style={line width=2.0pt}, error mark options={line width=2.0pt, mark size=2.0pt, rotate=90}]
 table[row sep=crcr, y error plus index=2, y error minus index=3]{%
2	11.9199382022472	0.040786224093029	0.040786224093029\\
3	17.1248881578947	0.428358549465577	0.428358549465577\\
4	21.5620074626866	1.83363140832695	1.83363140832695\\
};
\addlegendentry{\textbf{4 MIMO}}

\addplot [color=red, line width=2pt, mark size=2.0pt, mark=x, mark options={solid, red}]
 plot [error bars/.cd, y dir=both, y explicit, error bar style={line width=1.0pt}, error mark options={line width=1.0pt, mark size=3.0pt, rotate=90}]
 table[row sep=crcr, y error plus index=2, y error minus index=3]{%
2	5.54	0.082	0.082\\
3	8.733	   0.1 0.1\\
4	10.255	0.3	0.3\\
};
\addlegendentry{\textbf{8 MIMO}}



\end{axis}
\end{tikzpicture}%
    \end{subfigure}
        \vspace{-15pt}
    \caption{SE, EE of \name vs baselines}
    \label{fig:PESEfig}
    \end{minipage}
    \begin{minipage}[b][][b]{0.27\textwidth}
    \begin{center}
    \centering
\resizebox{\columnwidth}{!}{%
\begin{tabular}{|c|c|c|c|c|} 
 \hline
 Architec- &\# users & Good- & Power & RF  \\ [0.5ex] 
 -ture & &-put & (mW) & BW \\ [0.5ex] 
     & & (mbps) &  &  (MHz) \\ [0.5ex] 
 \hline
  & 2 & $\sim24$ & 562 & 10\\ 
  \name & 3 & $\sim36$ & 662 & 10 \\ 
  & 4 & 47$\pm$1.3 & 762 & 10\\ 
 \hline
  & 2 & $\sim24$ &  554 & 20\\
 OFDMA &  3 &$\sim36$  & 654 & 30  \\
  & 4 & $\sim48$ & 754 & 40 \\
 \hline
  & 2 & $\sim24$ & 4064  & 10\\
 8 Ant. & 3 & $\sim46$ & 4064 & 10 \\
 MIMO & 4 & $\sim48$ & 4064 & 10\\
 \hline
  & 2 & $\sim24$ & 2032  & 10\\
 4 Ant. & 3 & $33\pm2.3$ & 2032 & 10 \\
 MIMO & 4 & $39\pm4.3$ & 2032 & 10\\
 \hline
\end{tabular}}
\caption{Power consumption and Average $\pm$ Std. deviation (where relevant) of Goodput measurements  across $10$ testing positions in office setting for \name and baselines}
\label{fig:PESEtable}
\end{center}
    \end{minipage}
\end{figure}


\begin{figure*}
    \begin{subfigure}[t]{0.24\textwidth}
    \centering
    \setlength{\figureheight}{0.6\linewidth}
    \setlength{\figurewidth}{0.85\linewidth}
    \input{results/8ant_tput}    
    \caption{\name achieves target goodput}
    \label{fig:8ant_gput}
    \end{subfigure}
    \begin{subfigure}[t]{0.24\textwidth}
    \centering
    \setlength{\figureheight}{0.6\linewidth}
    \setlength{\figurewidth}{0.85\linewidth}
    \input{results/8ant_multipos_sajjad}    
    \caption{\name achieves target SINR}
    \label{fig:8ant_sint}
    \end{subfigure}
    \begin{subfigure}[t]{0.24\textwidth}
    \centering
    \setlength{\figureheight}{0.6\linewidth}
    \setlength{\figurewidth}{0.85\linewidth}
%
%
\begin{tikzpicture}

\begin{axis}[%
width=0.951\figurewidth,
height=\figureheight,
at={(0\figurewidth,0\figureheight)},
scale only axis,
xmin=3,
xmax=9,
xlabel style={font=\color{white!15!black}},
xlabel={\textbf{Number of Antennas}},
ymin=0,
ymax=1.4,
ytick={  0, 0.2, 0.4, 0.6, 0.8,   1, 1.2, 1.4},
ylabel style={font=\color{white!15!black}},
ylabel={\textbf{Capacity ratio}},
axis background/.style={fill=white},
xmajorgrids,
xminorgrids,
ymajorgrids,
yminorgrids,
xlabel style={font=\scriptsize},ylabel style={font=\scriptsize},legend style={font=\scriptsize},ticklabel style={font=\scriptsize},
ylabel shift={-6pt},
xlabel shift={-6pt}
]
\addplot [color=blue, line width=1.5pt, mark size=4.0pt, mark=x, mark options={solid, blue}, forget plot]
 plot [error bars/.cd, y dir=both, y explicit, error bar style={line width=1.5pt}, error mark options={line width=1.5pt, mark size=3.0pt, rotate=90}]
 table[row sep=crcr, y error plus index=2, y error minus index=3]{%
8	1.02310375269312	0.126452814832997	0.126452814832997\\
6	0.818397824946253	0.172638618150189	0.172638618150189\\
4	0.712117137593141	0.237987027798814	0.237987027798814\\
};

\addplot [color=black, dotted, line width=2.0pt, forget plot]
  table[row sep=crcr]{%
3 1\\
9 1\\
};
\node[draw=none] at (6, 1.2) {\scriptsize \textbf{OFDMA Capacity, 200 Mbps}};

\end{axis}
\end{tikzpicture}%
    \caption{\name attains capacity}
    \label{fig:8ant_cap}
    \end{subfigure}
    \begin{subfigure}[t]{0.24\textwidth}
    \centering
    \setlength{\figureheight}{0.6\linewidth}
    \setlength{\figurewidth}{0.85\linewidth}
    \input{results/trace_sim}
    \caption{Trace level MIMO comparisons}
    \label{fig:trace_lvl}
    \end{subfigure}
    \caption{Experiment CDFs for 100 data points collected across 10 different $4$ user configurations with (a) shwoing Goodput and (b) SINR increase as we increase \# antennas in \name (c) Plots the increase in average capacity due to improvements in SINR and shows \name hits OFDMA interference-free oracle capacity of 200 Mbps (d) Compares \name with $4,6,8$ MIMO configurations performed via trace-driven study
    }
    \vspace{-5pt}
\end{figure*}

\noindent \textbf{(b) SE and EE of \name, MIMO and OFDMA}

\textbf{Spectral Efficiency (SE)}: 
To calculate SE (bits per Hz), we devide the obtained goodput with RF bandwidth utilized to achieve the same.
The gold standard for SE would be to use $8$ Ant MIMO as baseline.
Basically even though number of users go from $2\to3\to4$, they occupy same $10$ MHz band and by using $8$ antennas $8$ Antenna MIMO would guarantee max. possible throughput to be always achieved (which are $24,36,48$ Mbps respectively).
\name achieves very close goodputs to $8$ Ant MIMO, by using $8$ antennas but keeping to user proportionate power consumption from $4$ virtual RF chains, as shown in Fig. \ref{fig:PESEfig} (a) and Fig. \ref{fig:PESEtable}.
OFDMA achieves similar goodputs, but at cost of higher spectrum requirements and hence OFDMA SE remains constant.
$4$ Ant MIMO uses $4$ physical RF chains and doesn't enjoy the added benefits of more antennas than number of users, hence for $4$ users the goodput is reduced and has a considerably high standard deviation as well due to increased bit errors.



\textbf{Energy Efficiency (EE)}: EE is calculated in bits per joule by dividing the goodput with power consumption in watts (Fig. \ref{fig:PESEfig} + Fig. \ref{fig:PESEtable}).
For EE, the gold standard is single antenna OFDMA, which consumes $354$ mW RF power from operation of MAX2829 RF transceiver used in WARP to amplify, filter and downconvert\cite{MAX2829}.
ADC power varies linearly with the sampling frequency \cite{barati2020energy,yan2019performance}, which is true for the AD9963 ADC used for WARP, as evident from the datasheet \cite{AD9963}, and it consumes $100$ mW per $10$ MHz sampled spectrum.
Hence to obtain the $24,36,48$ Mbps goodput, OFDMA spends just $354+200, 354+300, 354+400$ mW power, as tabulated in Fig.\ref{fig:PESEtable}.
\name achieves nearly the same throughput, but with just $8$ mW extra power ($562,662,762$ mW respectively for $2,3,4$ users), since \name uses 8 RF switches connected to $8$ antennas, with $1$ mW active power draw of $1$ HMC197BE switch of \name.
From the MAX2829 datasheet, we can see that 4 antenna MIMO would require RF power of 1632 mW (4$\times$408 mW) power when operating in synchronous MIMO mode.
This gets added with $4*100$ mW power for sampling $10$ MHz across the $4$ ADCs, and hence total power for 4 Ant MIMO is $2032$ mW.
We assume $2032*2 = 4064$ mW power for the trace driven study of $8$ antenna MIMO.

\begin{figure*}[t]
    \begin{subfigure}[t]{0.23\textwidth}
        \centering
        \setlength{\figureheight}{0.5\linewidth}
        \setlength{\figurewidth}{0.85\linewidth}
        \input{results/streams}
        \caption{$4$ Ant. \name's Multi-stream}
        \label{fig:mult_stream}
    \end{subfigure}
    \begin{subfigure}[t]{0.24\textwidth}
    \centering
    \setlength{\figureheight}{0.5\linewidth}
    \setlength{\figurewidth}{0.85\linewidth}
    \input{results/mmse_comparison_sajjad}
    \caption{SINR CDF for 2, 3, 4 users}
    \label{fig:4ant_sinr}
    \end{subfigure}
    \begin{subfigure}[t]{0.24\textwidth}
    \centering
    \centering
    \setlength{\figureheight}{0.5\linewidth}
    \setlength{\figurewidth}{0.85\linewidth}
    \input{results/random_seq_sajjad}    
    \caption{Choice of Switching matrix $\mathbf{S}$}
    \label{fig:random_seq}
    \end{subfigure}
    \begin{subfigure}[t]{0.24\textwidth}
        \centering
    \setlength{\figureheight}{0.5\linewidth}
    \setlength{\figurewidth}{0.85\linewidth}
    \input{results/unsynch_sajjad}   
    \caption{Synch. vs Unsynch. users}
    \label{fig:unsynch}
    \end{subfigure}
    \caption{Multi-stream results, Physical vs Virtual RF chains performance comparison and Ablation studies}
    \vspace{-5pt}
    \label{fig:ablation_etc}
\end{figure*}
\noindent\textbf{(c) \name can increase number of antennas to meet the target throughput and SINR requirements} 
\name allows antennas to be added without requiring complicated analog networks and at minimal power. Thus, by increasing the number of antennas, \name can arbitrarily adjust to a given target throughput/SINR/capacity requirements.

To show how using multiple antennas leads to these impressive results, we vary the number of antennas by electronically turn off antennas in \name PCB to simultaneously collect data from $4\to 6 \to 8$ antenna configurations.
We plot goodput, SINR and capacity results as we vary the number of antennas in Fig. \ref{fig:8ant_gput}, \ref{fig:8ant_sint} and \ref{fig:8ant_cap}.
Using antennas allows \name to get average throughput as $47$ Mbps very close to $48$ Mbps oracle throughput, also depicted in Fig. \ref{fig:PESEtable}, and SINR$>10$ dB even in the worst case.
At some user configurations, using $8$ antennas get $>15$ dB SINR as well, evident from the SINR CDF (Fig. \ref{fig:8ant_sint}).
However, at other configurations, mainly when the users are placed closer to each other interference suppression takes way some of the beamforming gains and gives capacity lower than OFDMA.
However, the average capacity of \name using $8$ antennas is near the OFDMA levels and the standard deviation is indicated by errorbars in Fig. \ref{fig:8ant_cap}.

Unlike \name, traditional MIMO adds antennas at cost of increased RF chains and hence increased power consumption.
We compare how adding more antennas in MIMO contrasts with \name, by doing a trace driven study, since WARPv3 has support only for $4$ physical RF chains. 
We collect first $4$ antennas channels at a time and then the next $8$ antennas and then stitch together the channel estimates in post-processing.
We observe that \name even outperforms a $6$ antenna MIMO in terms of SINR (Fig. \ref{fig:trace_lvl}), and comes $\sim3$ dB close to $8$ antenna MIMO at median performance.
As seen, both \name and $6,8$ antenna MIMO also meet 10 dB SINR requirements but the latter two approaches are not user-proportionate unlike \name, hence are suboptimal when EE is considered.

\textbf{(d) Using \name for multi-stream transmissions}:
In addition to multi-user, \name can also enable multi-stream transmission to increase throughput of a single user.
To test this setting, we synchronize clocks of $4$ USRP SBX daughterboards to act as a $4$ stream transmitting access point, and keep the antennas $\lambda/2$ close to each other, and receive via \name PCB equipped with $4$ antennas.
This emulates scenario of an access point transmitting to a smartphone, and hence we reduce number of antennas in \name.
We plot the goodput CDFs in Fig. \ref{fig:mult_stream}, and as we can see 4 ant \name is able to get up to $3$ independent streams very robustly, whereas $4$ streams transmission is not robust as is expected from a $4$ antenna \name. However, this motivates \name usecase to increase adoption of MIMO in smartphones, and reduce the burden on smartphone battery to enable such multi-stream transmissions.

\textbf{(e) Ablation studies}
The main design concepts of \name are creation of virtual RF chains from the spreaded bandwidth, BABF approach to beamform to one user per virtual chain.
We also test if interference cancellation performance depends on users being synchronized or not.

\textbf{(e.i) Physical vs Virtual RF chains}: 
To compare the physical and virtual RF chains in a fair manner, we do not turn on multiple antennas per virtual RF chain and restrict to one antenna per virtual chain.
We use same $4$ antennas connected to the $4$ physical RF chains, or \name PCB antenna ports by an external switch network (separate from \name's switches).
This switching network allows us to use the same antennas for physical vs virtual RF chain comaprison since now the channel observed across the chains would be same.
We collect the $2,3,4$ users interfering traces from this setup, and evaluate the average SINR across the users to evaluate the interference cancellation capacity of virtual vs physical RF chains.
Because the insertion losses from the switches are minimal ($\sim 0.5$ dB) and the PCB is designed with optimal Wilkinson networks at $2.4$ GHz, we see that the SINR metrics remain same (Fig. \ref{fig:4ant_sinr}) for both physical and virtual RF chains. 
This validates the \name PCB design and implementation.

\textbf{(e.ii) Why BABF?}: Using BABF approach, \name can use to beamform towards one user per virtual RF chain, and this enables a well-conditioned equivalent channel matrix $\mathbf{H}\mathbf{S}$ which is diagonal heavy.
Hence, as a consequence, the digital beamforming block which inverts the wireless channel works more efficiently and the final SINRs so obtained are much better.
These results are quantified as shown in Fig. \ref{fig:random_seq}, where for one $4$ user configuration we evaluate the final SINRs by choosing $\mathbf{S}$ randomly, and via BABF, and we repeat this 100 times to plot the CDF.
BABF always works better than any random choice of $\mathbf{S}$, and on an average (median line) it works $10$ dB better.
This justifies the design choice of adopting BABF approach of computing $\mathbf{S}$.

\textbf{(e.iii) Synchronized vs Unsynchronized Users}: \name performs a spatial interference suppression operation, hence it is oblivious to user synchronization.
We conduct experiments where we synchronize the $4$ users to share the same clock, and repeat it without synchronization.
We observe similar performance as expected, shown in Fig. \ref{fig:unsynch}.
However, \name needs at least loosely synchronized such that their LTS's do not collide to enable channel estimation. This is guaranteed by modern cellular protocols, and does not require users to share the full clocks.

\section{Discussion and Limitations}\label{sec:conlusion}

\name hardware experiments are limited to $4$ users with $8$ antennas in uplink setting. In this section, we show how \name can generalize for downlink, and more number of antennas/users and wider bandwidths.
We also show simulations with $8$ users and $64$ antennas which match the specifications of modern base-stations \cite{huaweiwhite,ericonmmimo,yan2019performance,han2020energy}, and compare with Hybrid beamformers (HBF).
Then, we present a case study on possible power savings in a 5G NR base station with \name.
We conclude with how \name can scale to wider bandwidths, and possible improvements to \name.
\begin{figure}[t]
\includegraphics[width=0.45\textwidth]{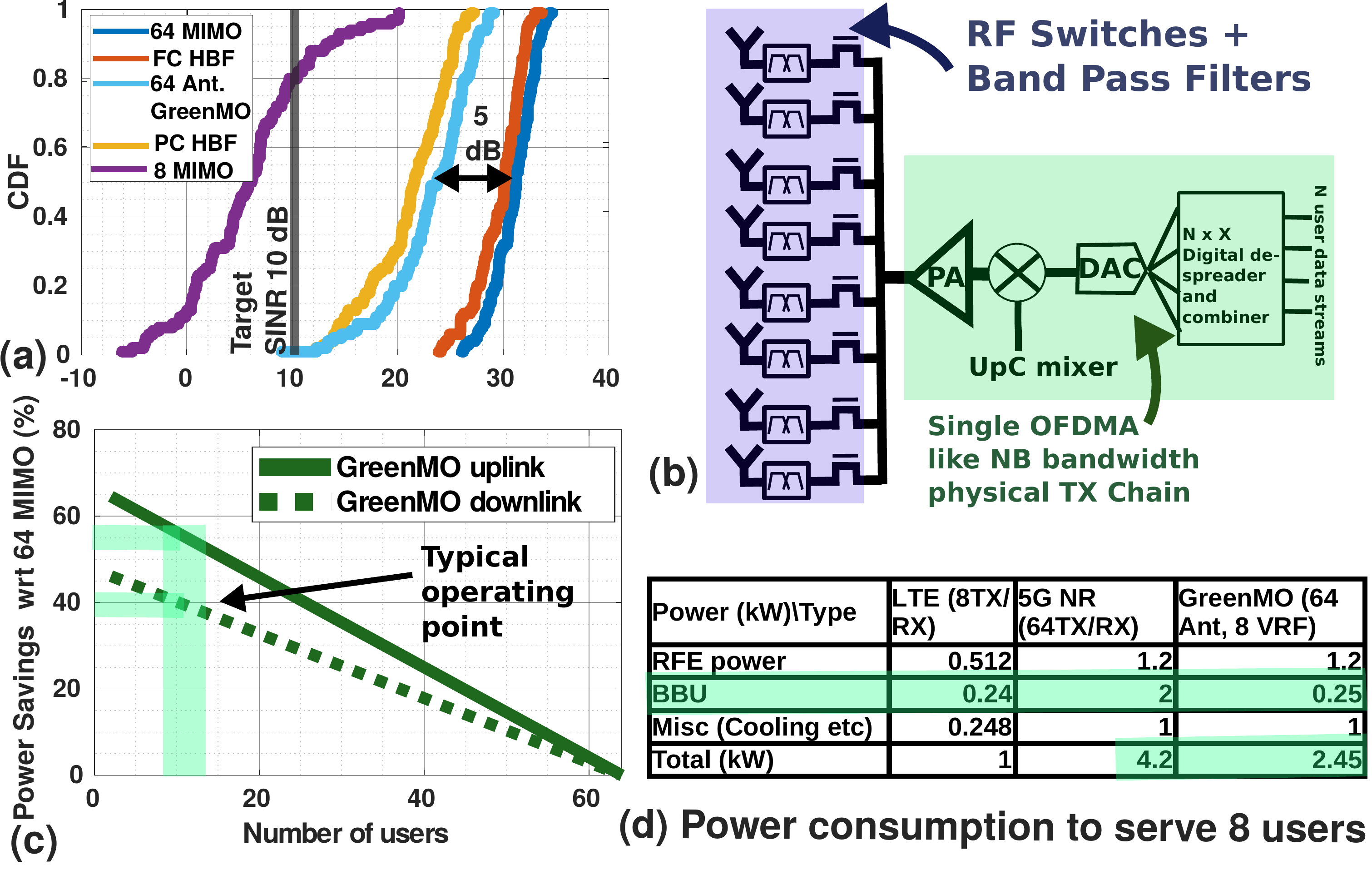}
\caption{(a) 8 user simulations, (b) possible \name downlink architecture and (c,d) Possible power savings in 5G NR base station with \name, at typical $8$ user operating point \name can achieve 40\% power savings in downlink and $57$\% in uplink}
\label{fig:discussion}
\end{figure}

\noindent \textbf{(a) $8$ user $64$ antenna simulations}: We perform simulations by placing $8$ users in similar environments to our evaluation setting, and modelling the multipath wireless channel via ray-tracing.
The simulations use identical transmit receive OFDM codes to our hardware experiments. 
The only difference is that noise is added artificially to maintain 15 dB SNR even in simulations, and the channel is applied via by calculating the distances from the simulation environment reflectors and applying appropriate time delays+amplitude weights. 
Hence, this simulation framework gives results which mimic the trends expected from evaluations.

Fig. \ref{fig:discussion} (a) shows the $64$ antenna simulations to serve $8$ users. This is considered a standard baseline with multi-user dense deployments, which on an average require serving $8$ streams concurrently \cite{yan2019performance}.
\name uses 64 antennas and creates 8 virtual RF chains to serve the 8 users, and almost always obtains >10dB SINR.
We can see even in simulations serving $8$ users with $8$ RF chain digital beamforming is infeasible.
We also show comparisons with partially ($8$ antennas per RF chain) and fully connected $64\to$all $8$ HBF. Our simulations assume perfect phase quantization in HBF, and still \name outperforms partially connected HBF. The reason is that unlike partially connected beamformers \name does not split the antenna array between the physical RF chains, since virtual RF chains by default behave like fully connected HBF, because of the many to many relationship between antennas and virtual RF chain. \name's performance comes about $5dB$ close at median level, to the full 64 chain DBF and fully connected HBF, even though using at least 64x lower power and 64x lower circuit elements.  

\noindent \textbf{(c) Downlink \name with filters and PA}:
We show a possible \name downlink architecture in Fig. \ref{fig:discussion} (b).
The downlink equivalency is motivated by the frequency domain picture of switching, since the DC harmonic would preserve the phases set by base band precoder and allow for downlink beamforming.
However, we need a filter right next to switch so that we do not transmit the other harmonic signals.
In downlink, we get additional power requirements for power amplifier (PA), and the PA energy consumption remains same for almost all possible architectures, like OFDMA, mMIMO, or even \name.
This is because PA power is EIRP driven, and OFDMA would need one stronger PA whereas mMIMO uses smaller PAs per antenna (but multiple of them) since it also gets antenna gain.
Infact, \name can also use smaller PAs per antenna after the splitter, but usually its better to have a single PA.
Hence, in downlink, \name would save the power consumption of baseband, but not contribute significantly in RF front end power savings as it is dominated by PA.
In case of uplink however, \name cuts the RF front end power to $1$ RF chain level.

\noindent \textbf{(b) How much power can \name save in a 5G NR base station?}
Major components of Base-station's radio energy consumption are PAs in RF front end (RFE), baseband units (BBU) and misc. cooling costs/other circuits like DC-DC converters \cite{han2020energy,auer2011much}.
Out of these components, baseband power depends on the ADCs/DACs operation and grows linearly with number of RF chains\cite{auer2011much,ha2013energy}.
So \name would help the base stations by making BBU power user-proportionate instead of consistently paying for all RF chains.

We list Huawei's commercial 5G NR \cite{huaweiwhite} and typical LTE macro-base station power consumption\cite{han2020energy,auer2011much} in Fig. \ref{fig:discussion} (d).
Hence, unlike LTE base stations, where RFE power is roughly 2x of BBU, in 5G NR BBU starts dominating the power consumption.
Also, the capacity gains from using $64$ chains is not $64\times$, but more like $8-16\times$ depending on channel conditions\cite{ericonmmimo}, and hence 5G base stations are being inefficient by paying $64\times$ power in BBUs.
Thus, by making BBU power user-proportionate, say for $8$ users it can be potentially reduced to LTE levels of $0.5$ kW using \name, could leading to power savings of at least $42$\%  (Fig. \ref{fig:discussion} (d)).
Note that, we have already shown that \name's 64 antenna 8 virtual RF chain simulations performance almost matches with full 64 MIMO in simulations.
Under bad channel conditions, or worst case if needed to serve $64$ users, \name is flexible and can simply mimic the full digital architecture to approx. same power consumption and not compromise on total performance.
In reality, \name would be slightly lower powered, since \name replaces multiple downconversion mixer with a single downconversion mixer, however the internal fraction of cost is not reported for 5G RFE \cite{huaweiwhite}.
We plot the potential power savings with number of users in Fig. \ref{fig:discussion} (c), and we can see how \name can make 5G base stations energy efficient by allowing them to respond to user load, in both uplink and downlink scenario.

\noindent \textbf{(d) Scaling to wider bandwidths}: We have evaluated $<4$ users, with bandwidth of $10$ MHz each. 
However, \name architecture can scale to wider-bandwidth with faster switches and higher sampling bandwidths. 
Photonic RF switches offer pico-second rise time\cite{ge2015ultra,xie2020sub,jiang2018ultra} which are 1000x faster than COTS solid state switches used in \name. 
Also, there are some expensive commercially available RF switches offering sub-ns rise times\cite{hmmc2027}. 
\name's architecture can create a new application scenario for faster switches as traditionally RF switches have been optimized only for improved isolation, or wideband operation, not for speed. 
On the other hand, ADCs capable of 1 GHz sampling bandwidth \cite{1gadcMaxim,1gadcTI} are commonplace now, which can support 50 virtual RF chains at 20 MHz sampling, typical bandwidth at sub-6 GHz. 
These 1 GHz sampling ADCs were thought to be useful only for mmwave frequencies where such wide over-the-air spectrum is available.
Hence, \name can enable these 1GHz ADCs to be useful even in sub-6 networks and unify the hardware needs of both sub-6 and mmWave bands

\textbf{(e) Handling neighbour band jammer and improved analog control over antennas}: Further, to improve \name uplink architecture, we need narrowband filters \cite{saw1,saw2} which would guard \name against a neighbour band jammer, since \name's analog spreader would spread the jammer as well into the band. Also, instead of using a single switch per antenna, using cascaded switch design would enable more granular control over antennas instead of $\{0,1\}$, hence further improve \name's results.

\textbf{(f) Other applications for \name}: Apart from base-station power reduction, \name can be used to enable energy-efficient MIMO in smart-phones to save battery, as motivated via $4$ antenna \name multi-stream results (Fig. \ref{fig:mult_stream}). \name can also be used for low-power MIMO in upcoming drone based APs, where it can extend the drone's battery life by reducing wireless power consumption \cite{buczek2022wireless}.

\section{Related Works}



\name presents for the first time, a MIMO architecture whose power is user-proportionate and can be scaled flexibly in response to any user load.
This is achieved via \name's approach of analog spreading digital de-spreading to create user-proportionate number of virtual RF chains.
In this section we will compare \name to other upcoming architectures, as well as some theoretical approaches.

Previously proposed MIMO interfaces which utilize a single RF chain \cite{tzeng2009cmos,johnson2020code,garg202028} form the closest of past work related to \name. These set of works also utilize a higher IF bandwidth than the users' transmitted bandwidth. However, instead of enabling creation of digital virtual RF chains, these works implement analog beamforming and utilize the higher bandwidth only to multiplex the outputs from these analog beamforming blocks.
Put more simply, these works either use IF bandwidth code domain \cite{tzeng2009cmos,johnson2020code} or different freq. bands in the higher IF bandwidth \cite{garg202028} to multiplex the outputs from a prior analog beamforming front-end.
The challenge with these works is that the analog beamforming component also needs to do beam-nulling, which is not very robust to wideband operation and phase shifter inaccuracies \cite{madani2021practical}.
In contrast, for \name architecture, the final interference cancellation block in \name is fully digital and hence \name can do wideband digital combining with arbitrary accuracy, which makes it stand out of the prior art on single-wire MIMO interfaces.

There are also a parallel set of works which explore parasitic antenna arrays to create artificial temporal wireless channel alterations \cite{alrabadi2010spatial,han2013mimo, ryu2015beam,hou2020low, sohn2016single, lee2016spatial, han2020characteristic,lee2019lambda}, as well as a body of works on Time Modulated Antenna Arrays (TMAA) \cite{jo2016demodulation15mhz,bogdan2019mimo,bogdan2019time,bogdan2020time,gonzalez2020wideband,bogdan2020lora,wang2016overview,gonzalez2018hybrid,chakraborty2021time,he2014space, maneiro2017time, bogdan2016experimental}.
These works either switching impedences of the parasitic antenna array, or modulate the antenna array with fast toggling RF switches.
These approaches are similar to how \name creates the frequency shifts using the RF switches.
For the parasitic array, there are known issues on the technique not scaling with number of antennas \cite{han2020characteristic}, and requires precision control over antenna impedance to generate the required orthogonal beams \cite{lee2016spatial}. 
The most notable of the TMAA set of works utilize a $4$ antenna array and a $50$ MHz switching speed to allow for increased diversity gains while decoding with a single RF chain using simple modulations like QPSK \cite{bogdan2019mimo,bogdan2019time,bogdan2020lora,bogdan2020time,bogdan2016experimental}. However, this has only be demonstrations are again limited to single user, unlike \name which enables multi-user spatial multiplexing instead of diversity.
Even for single user, there is no implementation extending these ideas to practical waveforms like OFDM. 
Thus to the best of our knowledge, \name remains the first demonstration of a `switched' antenna array which shows spatially multiplexed links with higher order constellations like QAM-16 and wideband OFDM transmissions.

Some papers have proposed to extend antenna arrays by stitching smaller arrays together via RF switches; mainly to improve spatial localization accuracy or bring diversity gains \cite{xie2018swan,gu2021tyrloc}. 
However, these works use static switches, whereas in \name switches are actively switching at baseband frequencies and the effect created by this fast switching is very different from the static on-off modelling in these papers.
There are other papers which target the uplink MIMO problem as well, with some of them requiring coordination from the users to do interference alignment \cite{gollakota2009interference,adib2013interference}, set random delays in transmissions to break channel correlations \cite{flores2016scalable}, or utilize distributed APs to serve multiple users \cite{wifimegamimo}. 
In contrast, \name utilizes just a single AP and demands no special coordination from the users, and hence the clients by default are COTS compliant.



\bibliographystyle{unsrt}
\bibliography{reference}

\end{document}